\documentclass[12pt,letterpaper,preprint]{aastex}
\newcounter{address}
\newlength{\figurewidth}
\usepackage{amssymb,amsmath,mathrsfs}
\newcommand{\project}[1]{\textsl{#1}}
\newcommand{\latin}[1]{\textit{#1}}
 \newcommand{\etal}{\latin{et~al}}
\newcommand{\setofall}[3]{\left\{{#1}\right\}_{#2}^{#3}}
 \newcommand{\allvn}{\Dvec_n}
 \newcommand{\allv}{\setofall{\allvn}{n=1}{N}}
 \newcommand{\allparsn}{{\omegavec_n}}
 \newcommand{\allpars}{\setofall{\allparsn}{n=1}{N}}
\newcommand{\hoggmatrix}[1]{\boldsymbol{#1}}
 \newcommand{\Dvec}{\hoggmatrix{D}}
 \newcommand{\alphavec}{\hoggmatrix{\alpha}}
 \newcommand{\omegavec}{\hoggmatrix{\omega}}
\newcommand{\est}[1]{\tilde{#1}}
\newcommand{\unit}[1]{\mathrm{#1}}
 \newcommand{\m}{\unit{m}}
 \newcommand{\s}{\unit{s}}
 \newcommand{\mps}{\m\,\s^{-1}}
 \renewcommand{\day}{\unit{d}}
\newcommand{\pomega}{\varpi}
\newcommand{\dd}{\mathrm{d}}
\newcommand{\like}{\mathscr{L}}
\newcommand{\documentname}{\textsl{Article}}
\newcommand{\equationname}{equation}
\newcommand{\sectionname}{Section}
\newcommand{\likealpha}{\mathscr{L}_{\alphavec}}
\newcommand{\falpha}{{f_{\alphavec}}}
\newcommand{\jupiter}{\mathrm{Jup}}

\begin{document}

\title{Inferring the eccentricity distribution}
\author{
  David~W.~Hogg\altaffilmark{\ref{CCPP},\ref{MPIA},\ref{email}},
  Adam~D.~Myers\altaffilmark{\ref{MPIA},\ref{UIUC}},
  Jo~Bovy\altaffilmark{\ref{CCPP}}
}

\setcounter{address}{1}
\altaffiltext{\theaddress}{\stepcounter{address}\label{CCPP} Center
  for Cosmology and Particle Physics, Department of Physics, New York
  University, 4 Washington Place, New York, NY 10003}
\altaffiltext{\theaddress}{\stepcounter{address}\label{MPIA}
  Max-Planck-Institut f\"ur Astronomie, K\"onigstuhl 17, D-69117
  Heidelberg, Germany}
\altaffiltext{\theaddress}{\stepcounter{address}\label{email} To whom
  correspondence should be addressed: \texttt{david.hogg@nyu.edu}}
\altaffiltext{\theaddress}{\stepcounter{address}\label{UIUC}
  Department of Astronomy, University of Illinois at Urbana-Champaign,
  Urbana IL 61801, USA}

\begin{abstract}
Standard maximum-likelihood estimators for binary-star and exoplanet
eccentricities are biased high, in the sense that the estimated
eccentricity tends to be larger than the true eccentricity.  As with
most non-trivial observables, a simple histogram of estimated
eccentricities is \emph{not} a good estimate of the true eccentricity
distribution.  Here we develop and test a hierarchical probabilistic
method for performing the relevant meta-analysis, that is, inferring
the true eccentricity distribution, taking as input the likelihood
functions for the individual-star eccentricities, or samplings of the
posterior probability distributions for the eccentricities (under a
given, uninformative prior).  The method is a simple implementation of
a hierarchical Bayesian model; it can also be seen as a kind of
heteroscedastic deconvolution.  It can be applied to any quantity
measured with finite precision---other orbital parameters, or indeed
any astronomical measurements of any kind, including magnitudes,
distances, or photometric redshifts---so long as the measurements
have been communicated as a likelihood function or a posterior
sampling.
\end{abstract}


\section{Introduction}

With rare exceptions, binary star and exoplanet science hinges not on
the specific value of any individual eccentricity (or mass or period),
but rather on the distribution, or the distribution as a function of
stellar properties or other parameters.  The goal of any statistical
study should be to determine the \emph{distribution} of the quantity
of interest---we will concentrate on orbital eccentricity for
specificity---that \emph{would} have been observed if the investigator
had extremely high signal-to-noise data and some (magical) method for
precise determination of all nuisance parameters, such as instrument
calibration and system inclination and so forth.  That is, the
investigator wants the observational-uncertainty-deconvolved
distribution of the quantity of interest.

One exciting development in the study of binaries and exoplanets is
that many groups are building probabilistic modeling software
(\citealt{ford05}; \citealt{gregory05}; \citealt{exofit}).  Rather
than fitting and returning a single set of parameters, These
probabilistic packages (approximately) \emph{sample} from the
posterior probability distribution under weak-prior assumptions.
These posterior samplings are much more useful than best-fit parameter
values, because they permit subsequent investigators to perform
probabilistic inference on the output without going back to the raw
radial-velocity data while still properly propagating uncertainties.
In this \documentname, we give an example of probabilistic
meta-analysis that becomes possible when the parameter outputs for
individual stars are probabilistic.

For any plausibly exoplanet- or binary-hosting star $n$, there are
parameters
\begin{eqnarray}\displaystyle\label{eq:exopars}
\omegavec_n &\equiv& \left( \kappa_n, T_n, \phi_n, e_n, \pomega_n \right)
\quad ,
\end{eqnarray}
where $\kappa_n$ is the velocity amplitude, $T_n$ is the period,
$\phi_n$ is some orbital phase or fiducial time, $e_n$ is the
eccentricity, and $\pomega_n$ is the longitude of perihelion.  Fitting
to these parameters is non-linear and unbiased estimators are rare.
For these reasons, maximum-likelihood (or, for Gaussian noise,
minimum-$\chi^2$) parameters are \emph{not} in general unbiased
estimators of the true parameters---maximum-likelihood estimators only
become unbiased in the limit of an infinite amount of data.  Almost
all single-point estimates, including maximum-\latin{a-posteriori} or
median-of-sampling parameters, are also biased in general
(\citealt{bias}).

Indeed, along these lines, it has been shown that the eccentricity
$e_n$ of any star $n$ has this property: The maximum-likelihood
estimate $\est{e}_n$ of the eccentricity is biased high; any histogram
of estimated eccentricities $\est{e}_n$ will have a mean (and
variance) that is higher than that of the true distribution of true
eccentricities $e_n$ (\citealt{shen08}).  These results---and the
incredible diversity of eccentricities observed in exoplanet
systems---motivate a concentration on the eccentricity distribution in
what follows.  Without the analysis methods we propose here, it is
possible that conclusions about the high eccentricities of exoplanet
systems might be over-stated or distorted.

There are three fundamental approaches to determination of the true
eccentricity dis\-tri\-bu\-tion---or true distribution of any
quantity---given noisy measurements $\est{e}_n$, where $n$ is an index
over instances (in this case, stars with binary or exoplanet
companions).  The first (and worst) is just to \emph{adopt} the
estimated values as good estimates of the true values and create a
histogram (or other density estimate) of the observed values
$\est{e}_n$.  Because these estimators are biased, and because they
are noisy, the distribution created in this way will have the wrong
mean and variance; it will be a strongly biased estimate of the true
distribution of true eccentricities $e_n$.  Furthermore, adding new
$\est{e}_n$ estimates from new stars $n$ will not decrease these
biases; there is no $\sqrt{N}$ improvement as new data are added.
There are suggestions of less-biased estimators for eccentricity
(\citealt{zakamska}), but anything unbiased in
eccentricity $e$ will still be biased for any nonlinear function of
$e$, and point estimates still have the property that the distribution
of point estimates will in general be different in \emph{variance}
from the true distribution, if only because of observational noise.

The second approach is to \emph{deconvolve} the distribution of
maximum-like\-li\-hood or best-fit $\est{e}_n$ values.  In this
approach the investigator recognizes that the distribution of
estimates $\est{e}_n$ is the true distribution convolved with the
uncertainty distribution, where that uncertainty can be described as
the probability of estimating $\est{e}_n$ when the true value is
$e_n$, or $p(\est{e}_n|e_n)$.  The investigator finds the distribution
of true values $e_n$ that, when it is convolved with the
uncertainties, produces the distribution of estimates $\est{e}_n$.
Done correctly, this will be performed in a forward-modeling of the
observed distribution, starting at the true distribution.  This method
is much more responsible, but when the investigator works at the
distribution level (that is, not at the individual-star level), the
investigator must assume things about the distribution of
uncertainties, equivalent to assuming that all the stars have the same
relationship $p(\est{e}_n|e_n)$ between the estimated and true values.
It is also a disadvantage that when performed na\"ively,
deconvolutions can be very sensitive to histogram binning (or,
equivalently, choices about the density estimation) and are unstable
to ``ringing'' and other issues coming from shot noise in the observed
distribution.

The third approach---and the approach taken here---is forward modeling
of the heteroscedastic observed data (or eccentricity estimates, if
these estimates are not single point estimates but rather posterior
samplings).  That is, the investigator makes a non-parametric (or
highly parameterized) model of the frequency distribution function
$\falpha(e)$ for the true $e_n$ values, and finds the best-fit values
of the distribution parameters $\alphavec$---the values of the
parameters $\alphavec$ that, after convolution with the (suitably
transformed) distributions in the nuisance parameters, explains best
the full set of eccentricity samplings.  This is also essentially
deconvolution, but it has the enormous advantage over na\"ive
deconvolution that it accounts for the fact that different stars have
different levels of (and functional forms for) uncertainty in the
nuisance parameters.  This forward-modeling approach is slowly being
adopted in astrophysics (an early proponent is \citealt{loredo04}); we
used a simple version of it to estimate the galaxy luminosity function
(\citealt{blanton03}) and the velocity distribution in the Galaxy disk
(\citealt{Bovy09a}), and we built a general tool for situations in
which distributions are smooth and observational uncertainties are
simple (\citealt{xd}).

Here we perform this forward modeling for a situation in which the
observational uncertainties are not simple (uncertainties are
asymmetric; measurements are biased) in an area (exoplanet
eccentricities) of great current scientific interest.  Everything that
follows is straightforwardly generalized to other parameters and other
kinds of systems.  For example, parallax-based stellar distances,
photometric redshifts, and faint-source fluxes also suffer from
systematic biases (\citealt{lutzkelker}; \citealt{photoz};
\citealt{hogg98}).  Scientific results based on these measurements
rely on the true distributions, not the distributions of (biased)
measurements, and in all of these cases, the objects of greatest
interest have measurements at relatively low precision or low
signal-to-noise.  Reliable scientific results can be obtained
nonetheless, though only by modeling the data; that is the fundamental
motivation for this work.

\section{Method}

There are $N$ stars $n$ ($1\leq n\leq N$), each of which has some
number $M_n$ of radial velocity measurements $v_{nj}$.  For each star
$n$, the set of measurements (data)
\begin{eqnarray}\displaystyle
\allvn & \equiv & \setofall{v_{nj}}{j=1}{M_n}
\end{eqnarray}
is modeled as being affected by a single companion.  We are not
explicitly considering multiple companion stars or planets at this
stage, although the generalization is straightforward.  The model is
\begin{eqnarray}\displaystyle\label{eq:model}
v_{nj} & = & V_n + g_n(t_{nj}) + E_{nj}
  \quad ,
\end{eqnarray}
where $V_n$ is an overall system velocity, the function $g(t_{nj})$ is
the radial velocity equation, and the $E_{nj}$ are noise contributions
drawn from a Gaussian of zero mean and variance
$[\sigma_{nj}^2+S_n^2]$, where $\sigma_{nj}^2$ is the uncertainty
variance for the $j$th observation of star $n$ and $S_n^2$ is a noise
variance from intrinsic stellar variability and other unmodeled
sources of noise.  The radial velocity equation $g_n(t)$ for star $n$
is parameterized by velocity amplitude $\kappa_n$, period $T_n$,
orbital phase $\phi_n$, eccentricity $e_n$, and longitude of perihelion
$\pomega_n$.  (five parameters).  This model of one star has 7
parameters ($V_n$, $S_n$, and five orbit parameters per star), which
we can think of as living in a list $\allparsn$, and the model of al
$N$ stars has $[7\,N]$ continuous parameters in a bigger list
$\allpars$.

The likelihood $\like_n$ for the seven parameters $\allparsn$ for star
$n$ is just the probability of the data $\allvn$ for star $n$ given
the parameters $\allparsn$ for star $n$
\begin{eqnarray}\displaystyle
\like_n & \equiv & p(\allvn|\allparsn)
 \nonumber\\
-2\,\ln\like_n & = & Q + \sum_{j=1}^{M_n} \ln(\sigma_{nj}^2 + S_n^2)
 +  \sum_{j=1}^{M_n} \frac{\left[V_n + g_{\omegavec_n}(t_{nj}) - v_{nj}\right]^2}%
                          {\sigma_{nj}^2 + S_n^2}
\quad ,
\end{eqnarray}
where $Q$ is some constant.  This looks like $\chi^2$ but is modified
for the jitter parameter $S_n^2$.

For each system $n$ we imagine that we have been provided (by the
exoplanet observing or fitting team, say) not the original data, but
just a $K$-element \emph{sample} from a posterior probability
distribution function (posterior PDF) created from the likelihood and
an uninformative prior PDF $p_0(\allparsn)$:
\begin{eqnarray}\displaystyle
p(\allparsn|\allvn) & = &\frac{1}{Z_n}\,p(\allvn|\allparsn)\,p_0(\allparsn)
\quad ,
\end{eqnarray}
where $Z_n$ is a normalization constant (for our purposes).  The prior
PDF $p_0(\allparsn)$ will be decided not by us but by the
exoplanet-fitter; we expect (need) it to be uninformative, for
example, flat in all parameters, or in their logarithms.  For each
star $n$ this sampling takes the form of a chain of $K$ samples $k$,
each of which is a set of 7 parameters $\allparsn_k$, such that the
distribution of the samples is consistent with a random draw from the
posterior PDF.

The total likelihood $\like$ for all the parameters of all the stars
$n$ is just the product of the individual-star likelihoods
\begin{eqnarray}\displaystyle
\like & \equiv & p(\allv|\allpars)
 \nonumber\\
      & = & \prod_{n=1}^N \like_n
\quad .
\end{eqnarray}
This product formulation of the total likelihood makes the implicit
assumption that the different star observations are independent; that
is, we are assuming that there are no likelihood covariances among the
parameters of \emph{different} systems $n$.  That assumption will be
at least weakly violated in any real survey of binaries or exoplanets,
because different observations will share hardware issues and
calibration information.

We want, however, not the likelihood for all the star and orbital
parameters but instead the likelihood $\likealpha$ for the parameters
$\alphavec$ of the eccentricity distribution $\falpha(e)$.  This
requires a slight re-thinking, because once we know the true
eccentricity distribution, \emph{that} is a better prior PDF to be
using than the uninformative prior PDF $p_0(\allparsn)$.  It is this
process---opening up the prior PDF to modeling and putting what we
want to infer in the place of the prior PDF from previous
inferences---that makes the method \emph{hierarchical}.  To make this
inference we must change variables and integrate out the
individual-star parameters which are---for our purposes---nuisance
parameters.  For the likelihood function $\likealpha$, this change of
variables and integration is
\begin{eqnarray}\displaystyle
\likealpha & \equiv & p(\allv|\alphavec)
\nonumber \\
\likealpha & = &
  \prod_{n=1}^N \int\dd\allparsn\,p(\allvn|\allparsn)\,p(\allparsn|\alphavec)
\nonumber \\
p(\allparsn|\alphavec) & \equiv &
  \frac{\falpha(e_n)\,p_0(\allparsn)}{p_0(e_n)}
\quad ,
\end{eqnarray}
where the integrals are over the $N$ 7-dimensional parameter spaces
and we have multiplied the uninfomative prior PDF $p_0(\allparsn)$ by
a ratio of the eccentricity distribution we want to infer to its
uninformative counterpart.  This is a \emph{marginalized likelihood}
(or ``marginal likelihood'') because we have inserted a prior PDF for
the nuisance parameters and integrated them out, but left the result
in the dimensions of likelihood (probability of data given
parameters).

In multiplying the prior PDF by $\falpha(e)/p_0(e)$ we have implicitly
assumed that the true distribution of parameters is \emph{separable};
that is, that both the uninformative prior PDF and the informative
prior PDF parameterized by $\alphavec$ can be written as a prior PDF
on the eccentricity $e$ multiplied by a prior PDF on the other
parameters.  This is not true in general, and is a limitation of this
formulation.  The limitation is not fundamental; $\falpha(e)$ can be
replaced with a multivariate distribution function in the general
case.

The $[7\,N]$-dimensional integral (or product of $N$ 7-dimensional
integrals) looks intimidating, but that integration is exactly the
capability that the sampling from each individual system $n$ provides
for us: Given a $K$-element sampling with elements $\allparsn_k$,
\begin{eqnarray}\displaystyle
\int \dd\allparsn\,p_0(\allparsn|\allvn)\,F(\allparsn) & \approx &
  \frac{1}{K}\,\sum_{k=1}^K F(\allparsn_k)
\quad ;
\end{eqnarray}
where $p_0()$ represents the posterior PDF under the uninformative
prior PDF upon which the sampling is based.  The point is that all
probability integrals can be approximated as sums over samples.  So
the sampling approximation to the marginalized likelihood for the
$\alphavec$ is just
\begin{eqnarray}\displaystyle
\likealpha & \approx &
  \prod_{n=1}^N \frac{1}{K}\,\sum_{k=1}^K \frac{\falpha(e_{nk})}{p_0(e_{nk})}
\quad ,\label{eq:like}
\end{eqnarray}
where all that is inside the sum is the \emph{ratio} between the
uninformative prior PDF (on which the sampling is based) and the new
prior PDF that we want to infer, and the $e_{nk}$ are the $K$ samples
of each $e_n$.  This sampling approximation to the likelihood
$\likealpha$ for the parameters $\alphavec$ can be optimized to obtain
a maximum-likelihood true eccentricity distribution, or it can be
multiplied by a prior PDF, normalized, and sampled to obtain a
posterior PDF sampling for the parameters $\alphavec$.  Either way, it
is the \emph{likelihood} that non-trivially enters into our inference.

The expression for the likelihood $\likealpha$ in
\equationname~(\ref{eq:like}) is an importance-sampling approximation
to the ratio of Bayes factors (integrals of the posterior probability
distribution over the nuisance parameters).  The comparison of
marginalized likelihoods of two models (between the default prior PDF
and the distribution parameterized by $\alphavec$ or between two
different values of the parameters $\alphavec$) is equivalent to a
marginalized Bayesian comparison between two models.  It is an
importance sampling because it uses a sampling but re-weights the
samples by the ratio of probabilities between the two models. The
usual caveats concerning importance sampling apply here as well: Since
the samples returned by the MCMC are generally not independent, the
importance-sampling approximation does not improve as $\sqrt{N}$ and
if the default prior PDF and the distribution parameterized by
$\alphavec$ are very different, the importance-sampling approximation
will be noisy. Therefore, we prefer the default prior PDF to be
uninformative.  We also need it to be uninformative to ensure that the
posterior PDF generated with $p_0(e)$ has support---and
samples---wherever the posterior PDF generated with $\falpha(e)$ is
significant.

From an inference perspective, it is more sensible to simultaneously
infer $\alphavec$ and all the $\allparsn$ for all the systems, and
perform the marginalization on the joint inference.  Here we use this
importance-sampling approximation because by assumption we do not have
the exoplanet data; we have only the $K$-element samplings from the
posterior PDFs (the posterior PDFs created with the uninformative
prior PDF).

All that remains is to choose functional forms for (parameterizations
of) the eccentricity distribution function $\falpha(e)$, and a prior
PDF on the parameters $\alphavec$ (if we want to perform sampling or
further marginalization, which we do).  There are many possible
choices here, and a true Bayesian doesn't choose but rather does them
all and includes them all in the output.  However, for specificity and
clarity, we consider only two forms for the eccentricity distribution.
The first is a step function with $M$ steps:
\begin{eqnarray}\displaystyle
\falpha(e) & \equiv &
  \sum_{m=1}^M \exp(\alpha_m)\,{\textstyle s(e;\frac{m-1}{M},\frac{m}{M})}
\nonumber \\
s(x;L,H) & \equiv & \left\{\begin{array}{cl}
    0 & \quad\mbox{for}~x<L \\
    (H-L)^{-1} & \quad\mbox{for}~L\leq x \leq H \\
    0 & \quad\mbox{for}~H<x \\
  \end{array}\right.
\nonumber \\
\sum_{m=1}^M \exp\alpha_m & = & 1
\quad ,
\end{eqnarray}
where we have laboriously defined the step function as a mixture of
top-hats and given the normalization constraint on the elements
$\alpha_m$ of the parameter vector $\alphavec$.  For the prior PDF on
the $\alphavec$ we use
\begin{eqnarray}\displaystyle
p(\alphavec) &\propto& \delta(1 - \sum_{m=1}^M \exp\alpha_m)\,\exp(-\frac{1}{2}\,\epsilon\,\sum_{m=2}^M [\alpha_m - \alpha_{m-1}]^2)
\quad ,\label{eq:prior}
\end{eqnarray}
where the delta function ensures the normalization of $\falpha(e)$,
and $\epsilon$ is a control parameter that controls our expectation
that $\falpha(e)$ be smooth.  Of course this smoothness parameter
$\epsilon$---and the bin number $M$---should be learned along with
$\alphavec$, but as this is beyond our scope, we simply set
$\epsilon=2$ and $M=20$.  Empirically, this keeps the distributions
smooth when the data sets get small, but permits good freedom and
doesn't influence the results much for large data sets, as we show
below.  This is a simple smoothness prior (\citealt{smoothness}); more
sophisticated versions can employ a Gaussian process
(\citealt{Rasmussen06a}) on the $\alpha_m$---the prior in
\equationname~(\ref{eq:prior}) is a special case of this---and the
hyper-parameters (only $\epsilon$ in this case) controlling the
smoothness of the distribution can be marginalized out. This has the
advantage that the number of bins can be very large, but it has the
disadvantage that sampling highly correlated bin heights is
challenging (\citealt{murray2010a, murray2010b}).

The second form we consider for the eccentricity distribution is that
of the beta distribution
\begin{eqnarray}\displaystyle
\falpha(e) &=& \frac{\Gamma(a+b)}{\Gamma(a)\,\Gamma(b)}\,e^{[a-1]}\,(1-e)^{[b-1]}
\nonumber\\
\alphavec &\equiv& (a, b)
\quad ,
\end{eqnarray}
where $\Gamma(z)$ is the gamma function, and we are redefining the
parameter list $\alphavec$ to contain the beta distribution shape
parameters $a$ and $b$.  This distribution is defined on the interval
$0<e<1$ and has remarkable freedom with only two parameters.  We take
the prior PDF on the parameters $(a,b)$ to be flat in the allowed
region $a>0$ and $b>0$.  Technically this prior PDF is improper, but
the posterior PDF under any realistic data set is proper.

In either case---step function or beta distribution---when we have a
set of $J$ samples $\alphavec_j$ of the distribution-function
parameter vector that are effectively samples from the posterior
probability distribution, we can use that distribution of
distributions or else marginalize out the parameters.  The
marginalized distribution $\left<\falpha(e)\right>$ is given by
\begin{eqnarray}\displaystyle
\left<\falpha(e)\right> = \frac{1}{J}\,\sum_{j=1}^J \falpha_j(e)
\quad ,
\end{eqnarray}
where by ``$\falpha_j(e)$'' we mean the distribution $\falpha(e)$ made using
parameters $\alphavec_j$ of the $j$th posterior sample.

\section{Experiments}\label{sec:experiments}

We generated $N=400$ ersatz radial velocity data sets $n$, each of
which contained $M_n=30$ radial velocity measurements scattered
uniformly over a time baseline of 1000 days. These measurements were
created using the model for the radial velocity of the parent star of
a theoretical exoplanet (or companion star in a binary system)
described by \equationname~(\ref{eq:model}). We assumed reasonable
distributions for the governing parameters of the model, together with
some (trivial) simplifications:

Each ersatz parent or primary star was assumed to have a mass of
$M_n=1\,M_{\sun}$ and vanishing overall system velocity ($V_n=0$) in
the absence of its exoplanet (or stellar companion). Intrinsic radial
velocity and primary star mass were, though, left free in
fitting. Each ersatz exoplanet or companion star was assigned a mass
$m_n$ drawn from a distribution $p(m)\propto 1/m$ (flat in $\ln m$)
constrained to lie in $[0.1\,M_{\jupiter}]<m_n<[10\,M_{\jupiter}]$.
Companions were given orbital periods drawn from a distribution
$p(T)\propto 1/T$ constrained to lie in
$[2\,\day]<T<[2000\,\day]$. Phases $\phi_n$ and $\pomega_n$ were drawn
from uniform distributions $0<\phi<2\pi$, except for the inclination
$i_n$, which was drawn from a distribution flat in $\cos i$.  The
masses $M_n$ and $m_n$ and inclinations $i_n$ enter into the radial
velocity model through the $\kappa_n$:
\begin{eqnarray}\displaystyle
\kappa_n &=& \frac{[2\pi\,G]^{1/3}\,m_n\,\sin i_n}{T_n^{1/3}\,[M_n+m_n]^{2/3}\,[1-e_n^2]^{1/2}}
\quad .
\end{eqnarray}

The eccentricity $e_n$ for each ersatz companion was drawn from one of
two frequency distributions: The first is an intuitive distribution
\begin{eqnarray}\displaystyle
f(e)= \frac{1}{Z}\left[\frac{1}{[1+e]^4}-\frac{e}{2^4}\right]
\end{eqnarray}
(\citealt{shen08}), where $Z$ is a normalization constant; hereafter
we call this distribution ``ST4''.  The second is an unrealistic
straw-man designed to stress-test the inference methodology.  It is a
Gaussian with mean $e=0.3$ and variance $(0.05)^2$ (but set to zero
outside of the range $0<e<1$).

The error added to each ersatz measurement was drawn from a Gaussian
of known observational noise variance, but with every point assigned
its own individual (heteroscedastic) noise variance, uniformly
distributed (in variance) in the interval
$(\sqrt{10}~\mps)^2<\sigma_{nj}^2<(10~\mps)^2$.  For the ersatz
observations, the jitter parameters $S_n^2$ were set to vanish, but,
as with the system velocity and mass of the parent (or primary) star,
the jitter parameters $S_n^2$ were left free in the fitting.

We optimized, fit, and marginalized the radial velocity model to each
of the ersatz data sets using Metropolis-Hastings Markov Chain Monte
Carlo (MCMC) sampling.  For the MCMC we adopted a standard
uninformative prior PDF that is flat in $\ln\kappa_n$, flat in $\ln
T_n$, flat in $\phi_n$, flat in $\pomega_n$, and flat in $e_n$.  We
performed the sampling not in the naive parameter space
$(\ln\kappa_n,\ln T_n,\phi_n,e_n,\pomega_n)$ but in a better-behaved
sampling space of
\begin{eqnarray}\displaystyle
(\ln T_n, \kappa_n\,\cos[\phi_n+\pomega_n]
        , \kappa_n\,\sin[\phi_n+\pomega_n]
        , e_n\,\cos\pomega_n
        , e_n\,\sin\pomega_n)
\quad.
\end{eqnarray}
In this space, sampling is better behaved, but the ``natural prior''
PDF is not flat in $\ln\kappa_n$ or $e_n$, so a compensating prior PDF
(or Jacobian) must be multiplied in.  We confirmed that our sampler is
using the correct uninformative prior PDF by running it on empty data
sets; the resulting no-data samplings are samplings of the prior PDF.

\begin{figure}
\includegraphics[width=\figurewidth]{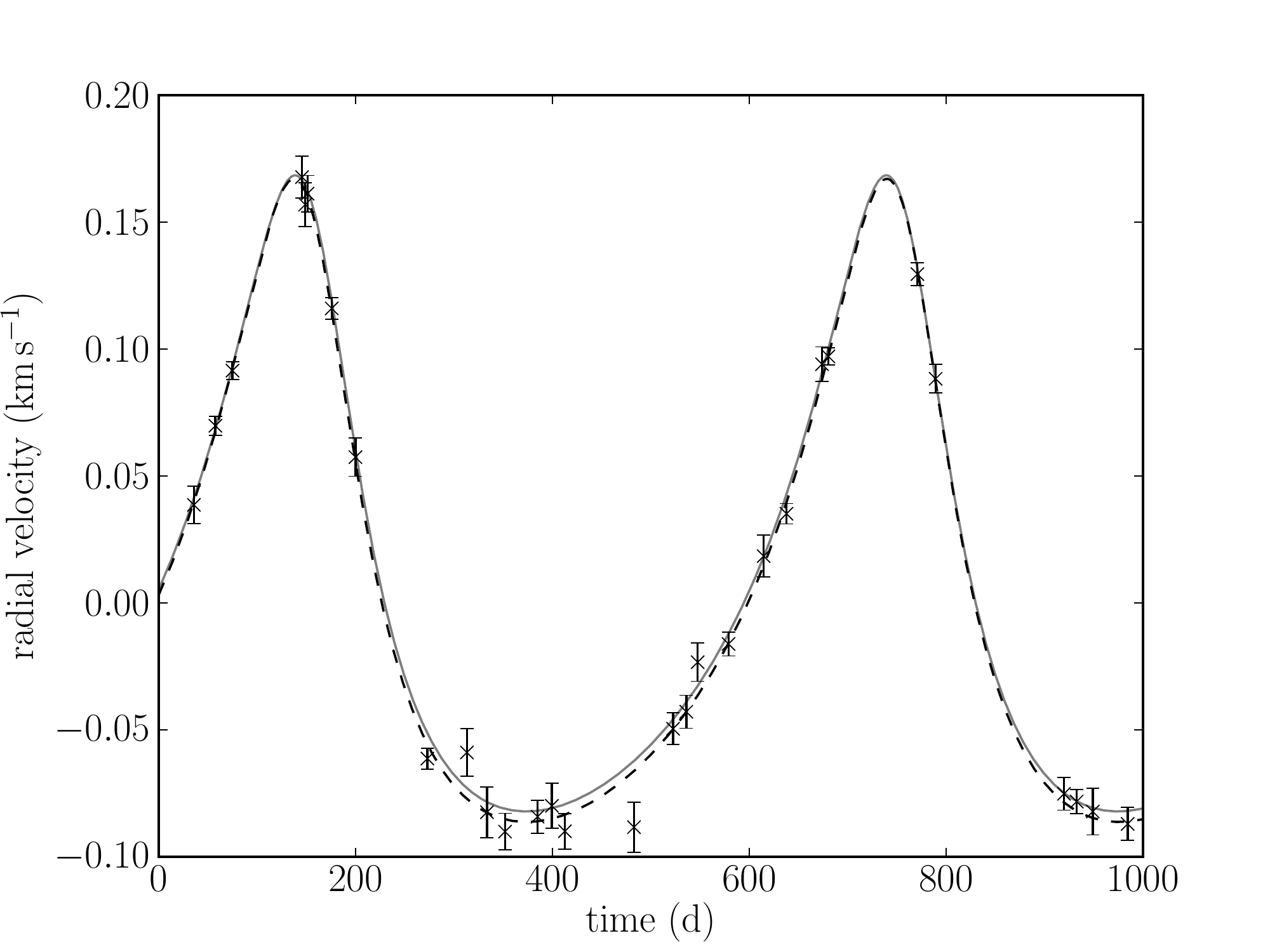}
\includegraphics[width=\figurewidth]{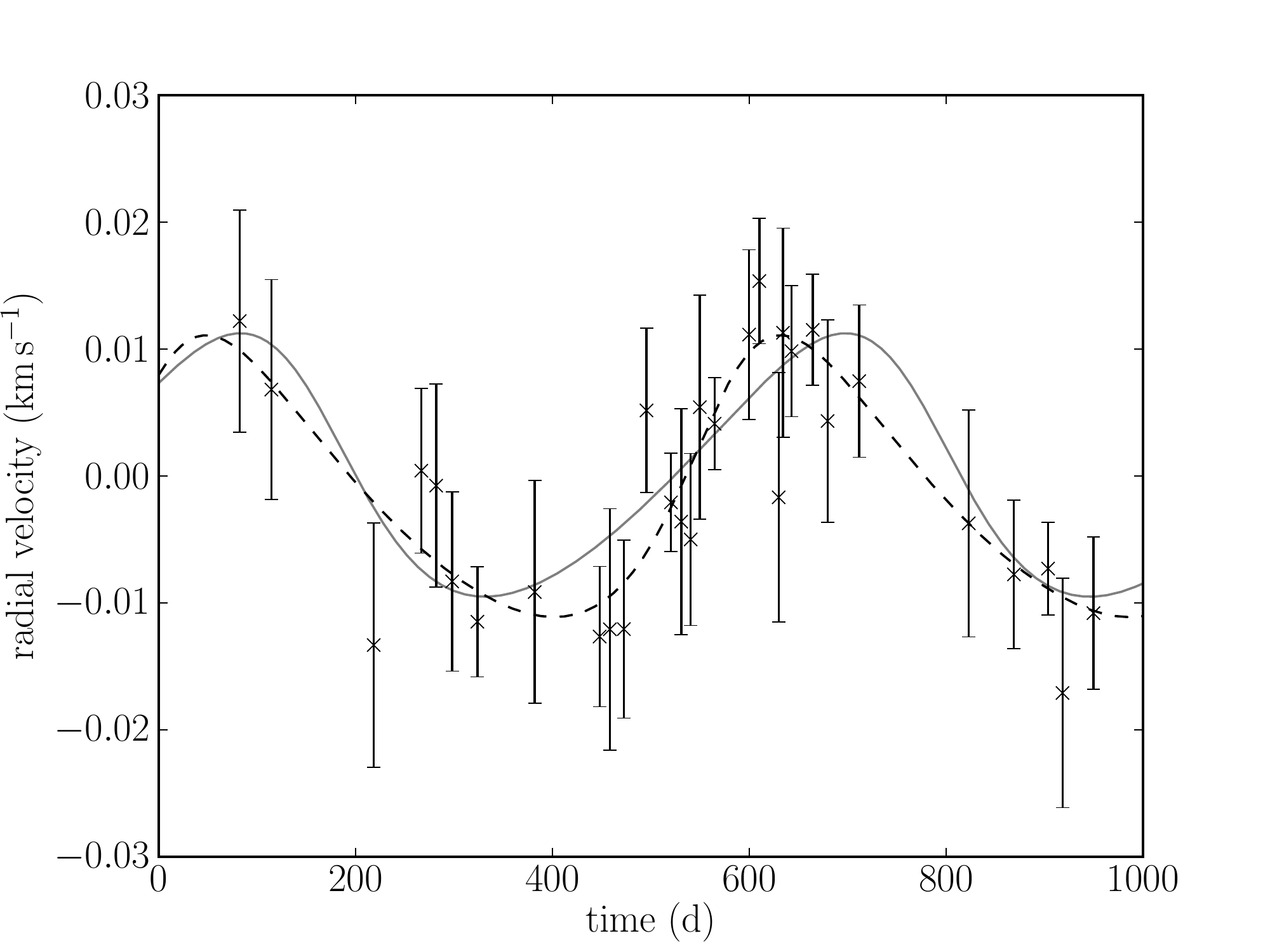}\\
\includegraphics[width=\figurewidth]{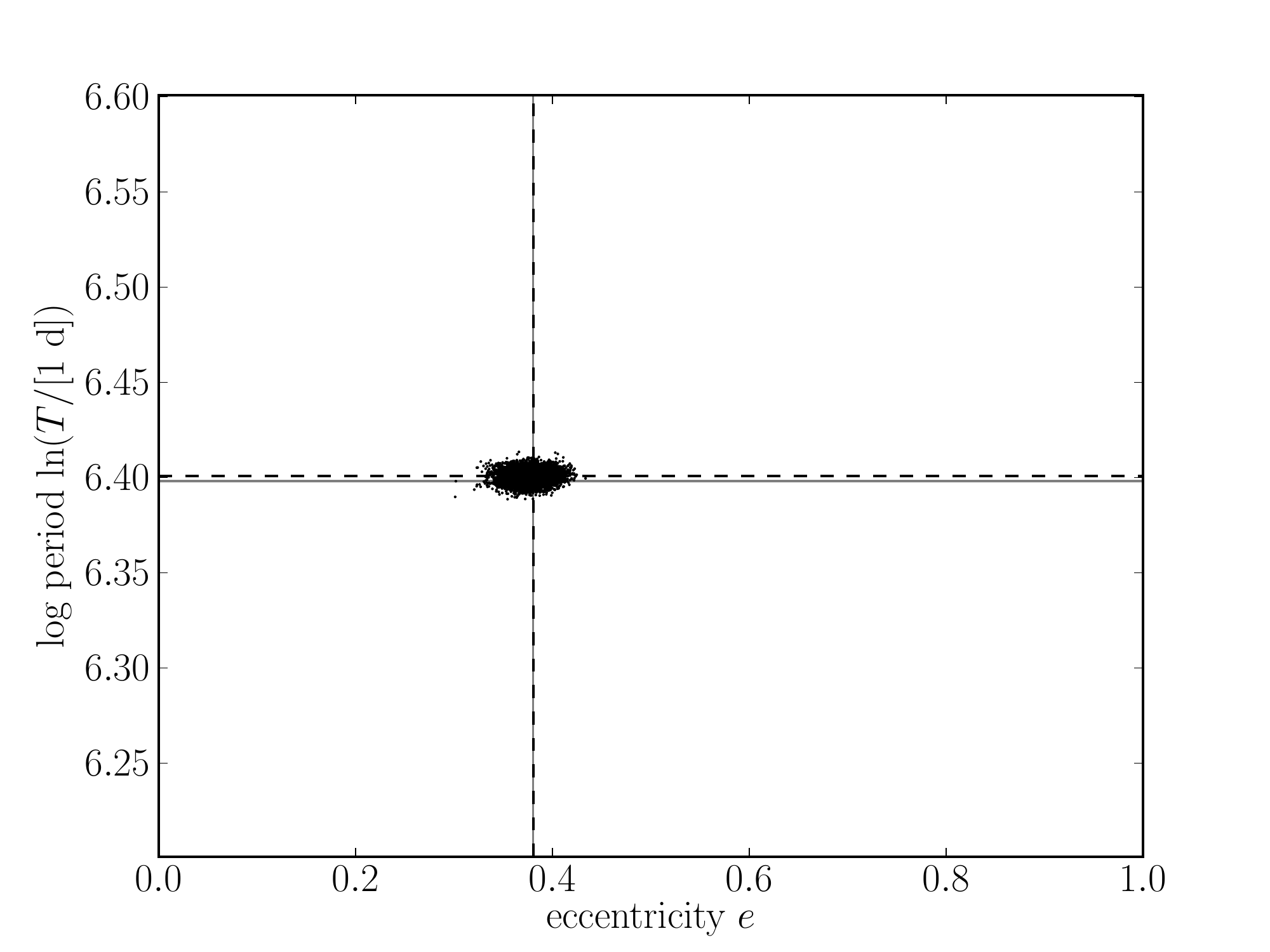}
\includegraphics[width=\figurewidth]{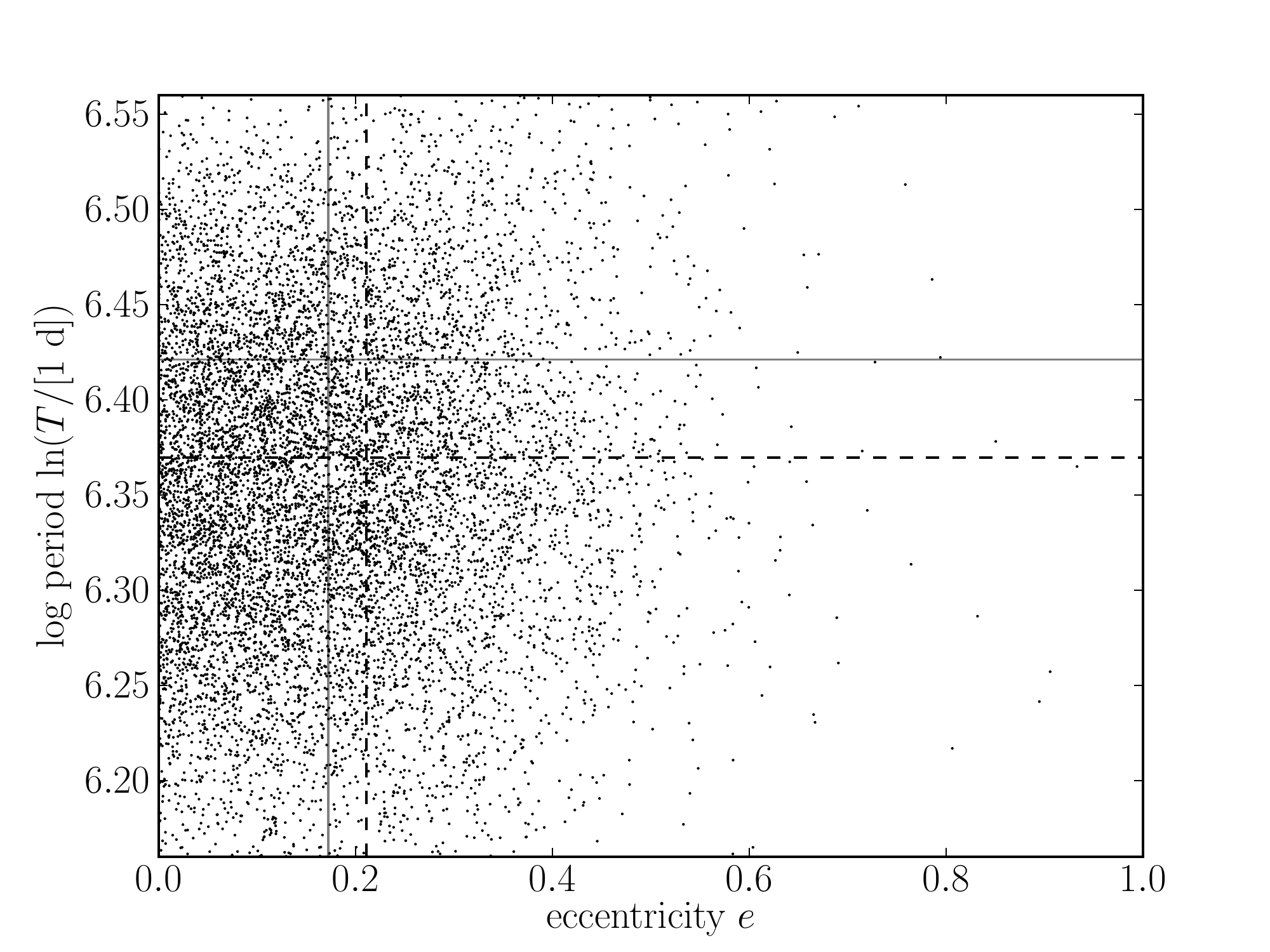}\\
\includegraphics[width=\figurewidth]{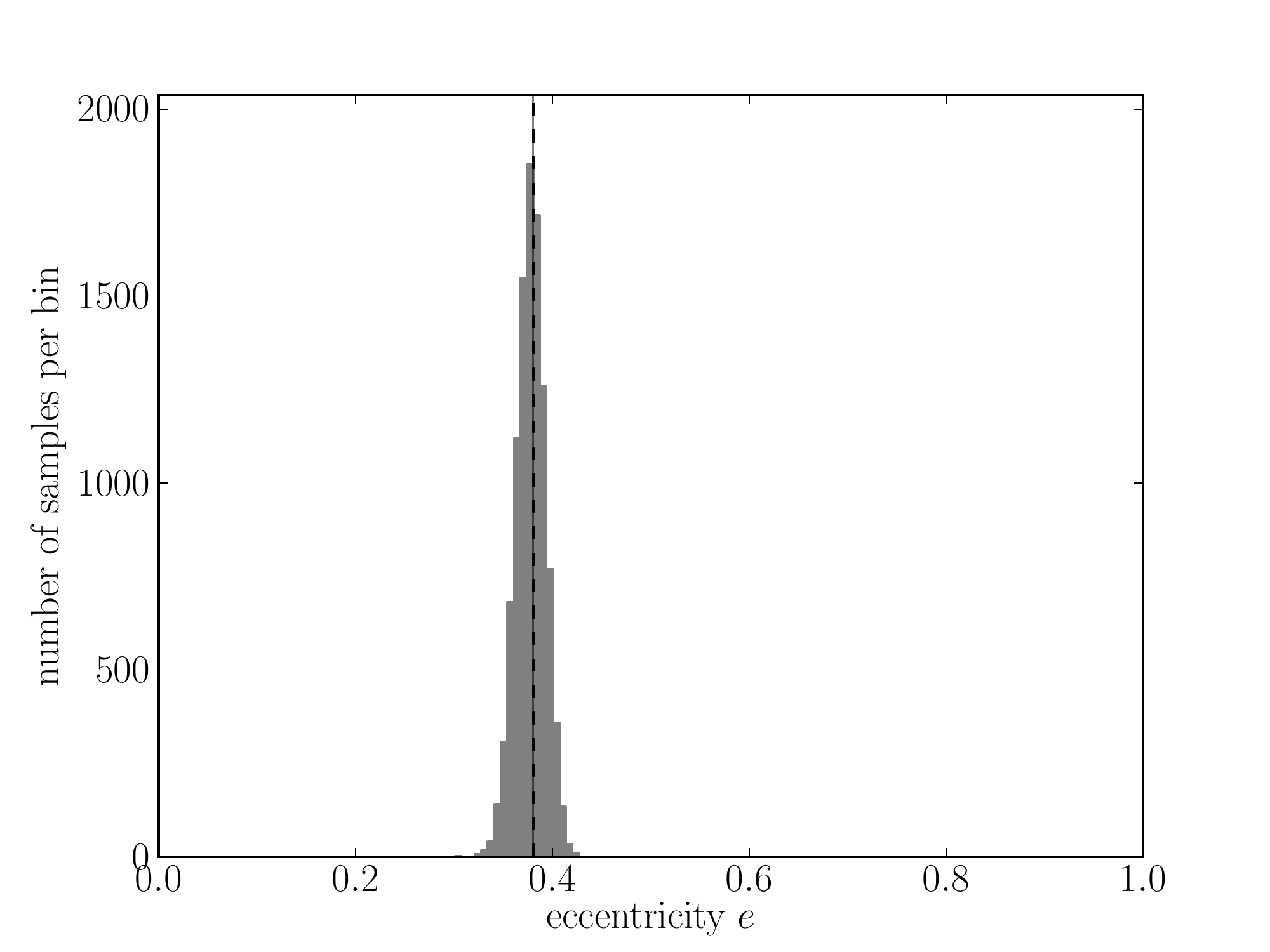}
\includegraphics[width=\figurewidth]{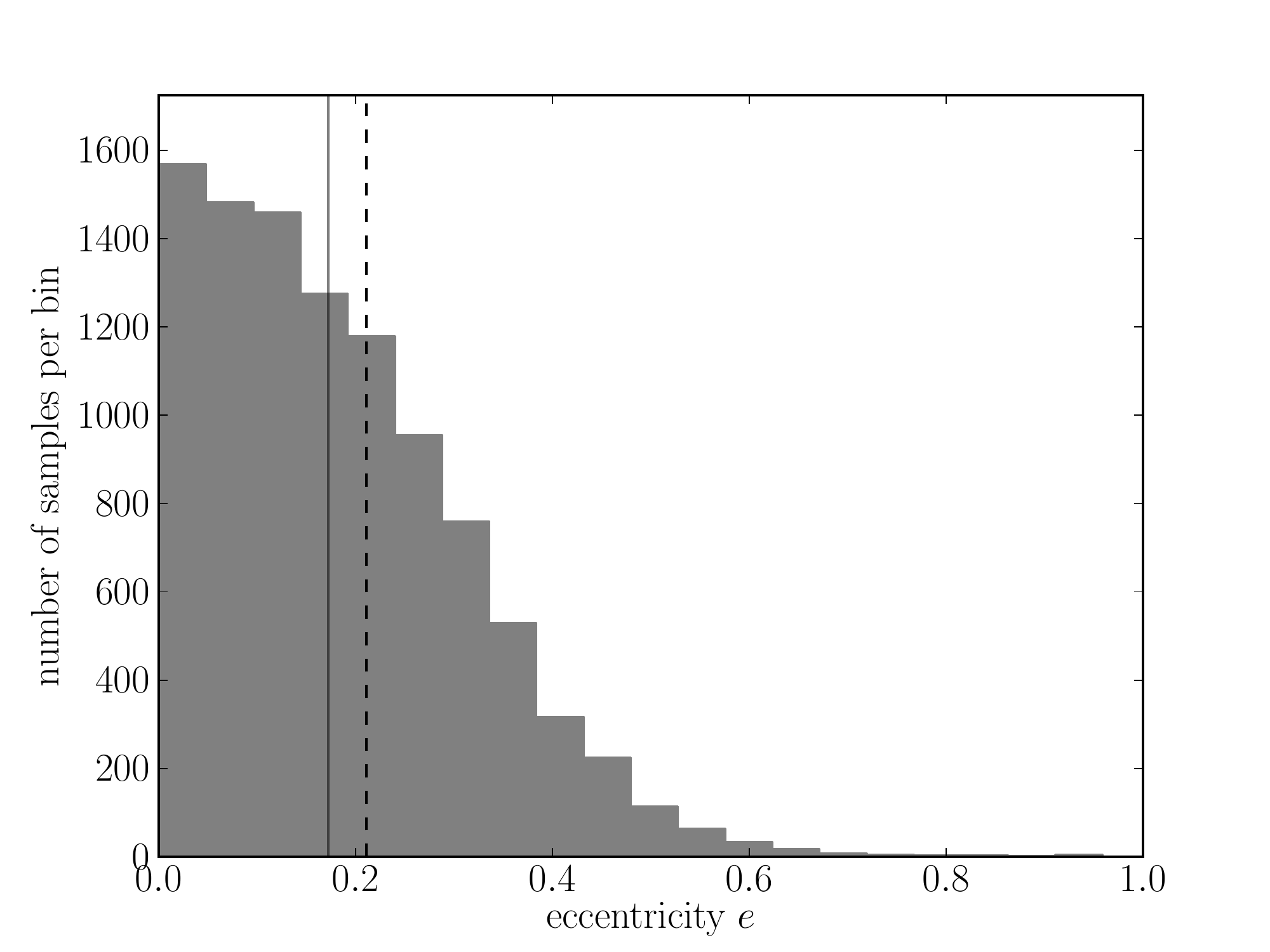}
\caption{Two example ersatz exoplanets, one with relatively high
  signal-to-noise and one with relatively low.  The top panels show
  the heteroscedastic radial velocity data with the true radial
  velocity model overplotted as a solid grey line, and the the
  maximum-likelihood (best-fit) radial velocity model overplotted as a
  dashed black line.  The middle and bottom panels show the
  distribution of periods and eccentricities in ten percent of the
  $K=10^5$ samples in the thinned chain from the MCMC.  Again, the
  solid grey line shows the true value, and the dashed black line
  shows the maximimum-likelihood value.\label{fig:exoplanet}}
\end{figure}

On each system $n$ we set parameter step-sizes (for a Gaussian
proposal distribution in the sampling space) such that acceptance
ratios were near $0.4$.  On each system $10^6$ links of MCMC were run,
checked for mixing (convergence), and thinned (subsampled uniformly)
to produce a set of $K=10^5$ nearly independent parameter samples
$\allparsn_k$---independence is however not required for the sampling
approximation of \equationname~(\ref{eq:like}) to work.  Radial
velocity curves for two ersatz exoplanets are shown in
\figurename~\ref{fig:exoplanet}, together with their MCMC samplings in
period and eccentricity.  For each system $n$ we search the chain for
the maximum-\latin{a-posteriori} parameters.  We also used a modified
(simulated annealing) MCMC sampling to find the maximum-likelihood
parameters $\est{\allparsn}$, one member of which is the ``best-fit''
eccentricity $\est{e}_n$.

At this point we have the full $N\times K$ sampling $e_{nk}$.  This
makes it possible to compute the marginalized likelihood of
\equationname~(\ref{eq:like}) for any parameters $\alphavec$.  Again,
we perform Metropolis-Hastings MCMC, but now in the space of
$\alphavec$ with the prior PDF of \equationname~(\ref{eq:prior}).  The
proposal distribution used in the MCMC was a small Gaussian
perturbation applied to every component of $\alphavec$, with Gaussian
variance chosen to keep the acceptance ratio near 0.4.  For each
experiment $10^4$ links of MCMC were run, and checked for mixing.
From this posterior sampling, we can obtain whatever results are
desired, the maximum-\latin{a-posteriori} distribution, the
marginalized (mean-of-posterior) distribution, a sampling of
distributions consistent with the data, or any quantile of the
eccentricity distribution.

In \figurename~\ref{fig:300} and \figurename~\ref{fig:30} we show the
results of four experiments, two with $N=300$ systems and two with
$N=30$, and two with the ST4 input eccentricity distribution, and two
with the Gaussian input eccentricity distribution.  We show the true
(input) eccentricities, the maximum-likelihood eccentricities, and the
inferred distribution.  In each case, we find---as expected---that the
marginalized (mean-of-posterior) inferred eccentricity distribution is
a far better description of the original input eccentricity
distribution than the naive distribution created by histogramming the
maximum-likelihood eccentricity estimates $\est{e}_n$.  The inferred
distribution captures well the smooth distribution that was used to
generate the ersatz data.  Furthermore, where the actual finite
sampling of true values departs (just by Poisson statistics) from the
smooth distribution, the inferred distribution even captures those
deviations, especially when there are $N=300$ systems
(\figurename~\ref{fig:300}).

\figurename~\ref{fig:300} and \figurename~\ref{fig:30} also show that
the mean of the eccentricity distribution is over-estimated when the
best-fit eccentricities $\est{e}_n$ are used to represent the
eccentricity distribution, and that the over-estimate does not
decrease as the number of objects increases.  However, the
marginalized inferred mean---the mean obtained by marginalizing over
samples---is a very good estimate of the true mean of the true
distribution. The same holds for all of the quantiles of the
distribution.

The sampling in $\alphavec$ provides full uncertainty information,
including the uncertainty in every component of $\alphavec$ and also
all of the component--component covariances.  In
\figurename~\ref{fig:300} and \figurename~\ref{fig:30} we show only a
superimposed sampling.  This conveys information about the uncertainty
distribution in each bin, but it doesn't display the full power of the
sampling for error analysis and propagation.

We repeat these experiments but now using the beta distribution for
$\falpha(e)$.  Once again we perform Metropolis-Hastings MCMC, but now
in the space of the beta-distribution shape parameters
$\alphavec\equiv (a,b)$.  Again the proposal distribution used in the
MCMC was a small Gaussian perturbation applied to the two components
of $\alphavec$, with Gaussian variance chosen to keep the acceptance
ratio near 0.4.  For each experiment $2\times10^3$ links of MCMC were
run, and checked for mixing.  We show the results of the
beta-distribution fitting in \figurename~\ref{fig:beta}.  It also
performs extremely well.

We obtained $K=10^5$ samples per star, but this is almost certainly
overkill.  In \figurename~\ref{fig:K50} we show how the results change
when this sampling is thinned down to just $K=50$ samples of the
posterior PDF.  The thinning was done uniformly, taking every 2000th
sample from the parent set of $10^5$.  The results are only slightly
worse with $K=50$; this speeds up the code by a factor of 2000, of
course.

\section{Discussion}

We have shown that proper inference of the eccentricity distribution
outperforms the naive approach of histogramming best-fit eccentricity
values.  Our inference proceeds by inserting the model eccentricity
distribution as a prior on the eccentricities, after which a good
eccentricity distribution is one that makes the data---the set of all
exoplanet observations---probable.  Because this method models the
distribution prior to observation, it is effectively a deconvolution
working on heteroscedastic data; it is a generalization to arbitrarily
non-Gaussian uncertainties of previous work in this area
(\citealt{xd}).  There is nothing crucial about eccentricity; the
method presented here can be applied to any problem in which the true
distribution $f(x)$ is desired and there are only noisy measurements.
As long as the measurements are presented as likelihood functions or
samplings under an uninformative prior, the function $f(x)$ can be
inferred as described here.

Though it is common for investigators to present as measured
distribution functions the histogram of estimated values, sometimes
even worse mistakes are made.  For example, it is sometimes
tempting for an investigator to see the output of the sampler as
providing a better estimate of the distribution function.  The
thinking is ``well, the object has some probability of being in each
of these eccentricity bins, so I will add a bit into each bin on its
behalf''.  This thinking is wrong: It convolves the error-convolved
distribution with the errors once again.  We will not cite specific
examples (to protect the guilty), but this is occasionally done and
always incorrect.

One limitation of this study is that we made a simplifying assumption
of a separable prior; that is, we imagined that there was one
eccentricity distribution for all of the stars in the sample.  In
reality, the eccentricity distribution will depend on parent star and
exoplanet parameters, and there will be no clean separation of the
eccentricity distribution.  There are two reactions to this.  The
first is to live with the result, understanding that the distribution
returned by the method is correct for the specific sample in hand,
marginalizing over all other parameters.  The second is to permit
non-separable distribution functions.  In this latter case,
hierarchical modeling becomes even more necessary. All of the
correlations between orbital parameters or dependences on the parent
star's properties can be modeled on the prior level and fit as we did
here. This situation is not substantially more complicated; it is just
that the terms in the sum in the likelihood of
\equationname~(\ref{eq:like}) becomes a function of multiple
parameters, and the parameterization of the function changes.

Another limitation of this work is that we only considered
single-planet systems.  This was in part because we do not have a
simple prior for generating realistic multiple-planet systems, but
also because there is nothing fundamental that changes if we switch to
multiple-planet systems.  We also permitted additional ``jitter'' in
the observations but did not in fact add jitter or any other kind of
additional noise or data outliers.  Again, addition of these
things---and proper modeling of them---changes the individual-object
likelihood functions and samplings, but does not change the procedure
by which the eccentricity distribution is inferred.

Finally, this study has some danger of ``garbage in, garbage out:'' we
generated data in accord with our general expectations, and fit it in
accord with those same expectations.  This is a limitation of all
studies on artificial data, of course.  However, we note that the
parameterized eccentricity distributions that we fit are mixtures of
step functions and beta distributions; neither of these are related
directly to---or even very appropriate for representing---either of
the distributions we used to generate the ersatz data.  Furthermore,
the ``uninformative'' prior PDF we used in the exoplanet samplings
were wrong, not just because it got the eccentricity prior PDF wrong
(the subject of this \documentname) but because they also got the
velocity amplitude $\kappa$ prior PDF wrong: The ersatz exoplanets
were generated with isotropic inclinations and a power-law
distribution of masses $m$; this does not generate a power-law
frequency distribution in $\kappa$.  That is, this method works well
even when the priors in the other parameters are substantially wrong.
And of course these prior PDFs can be inferred too, in a
generalization.

\begin{figure}
\includegraphics[width=\figurewidth]{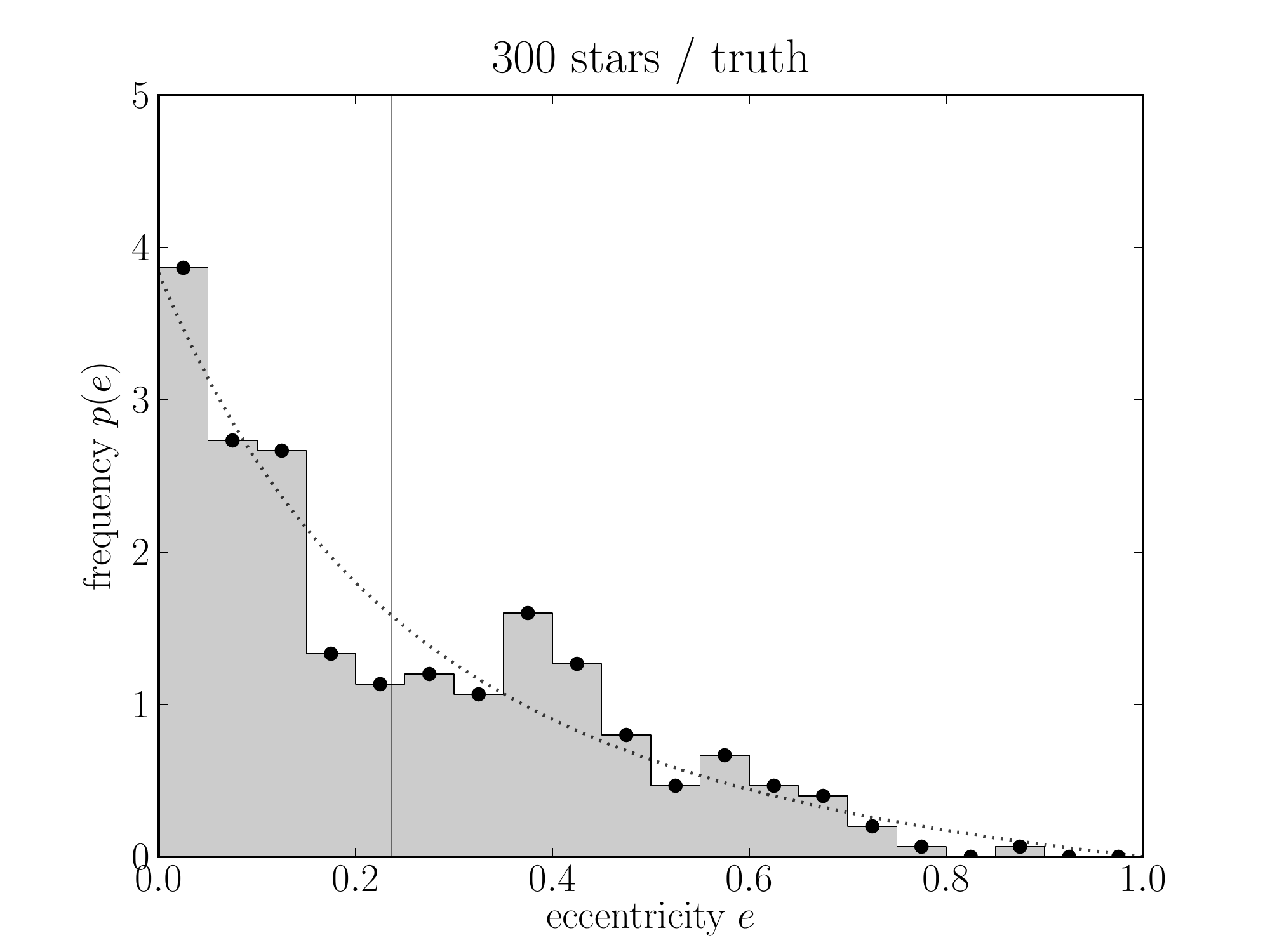}
\includegraphics[width=\figurewidth]{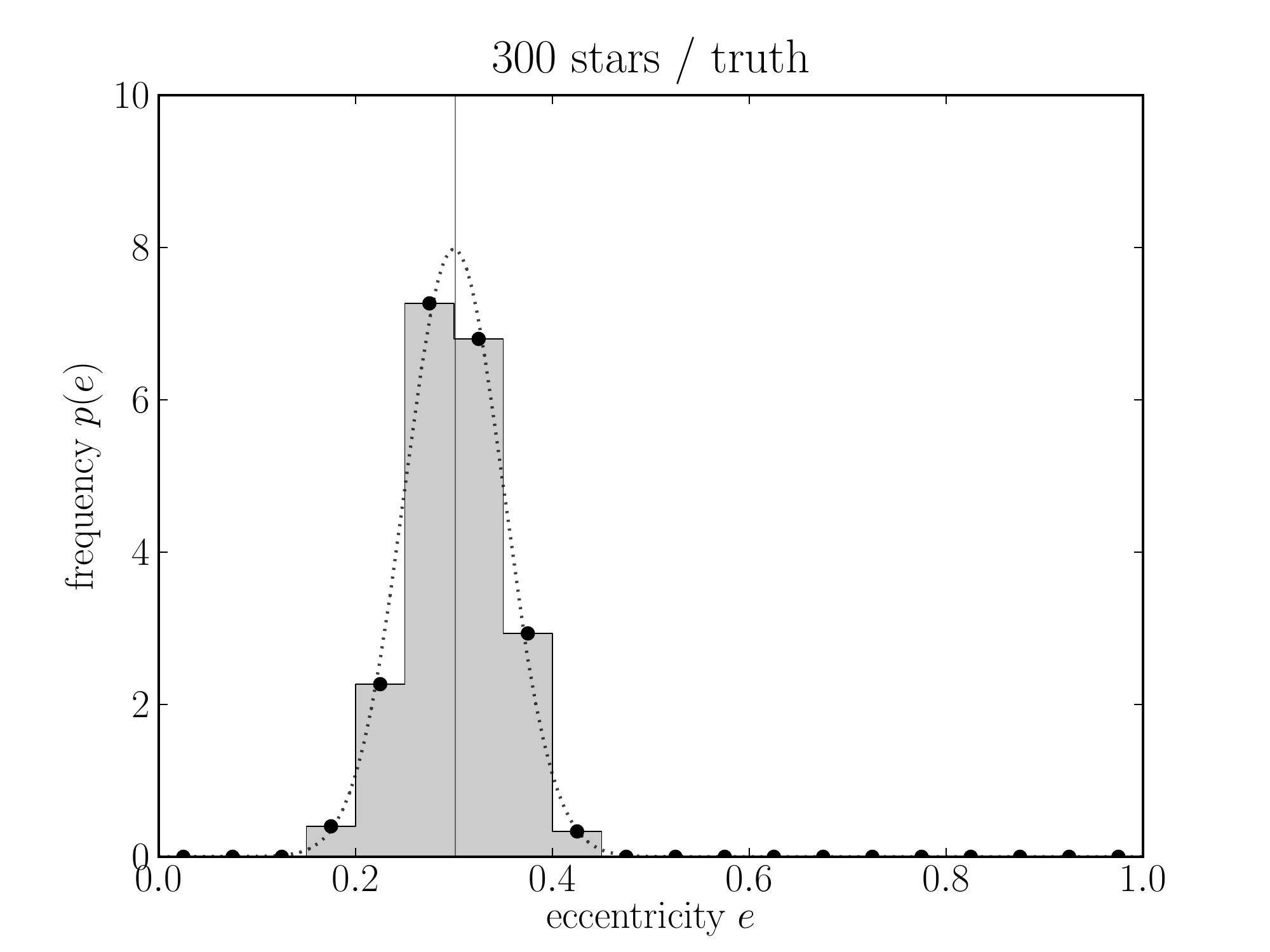}\\
\includegraphics[width=\figurewidth]{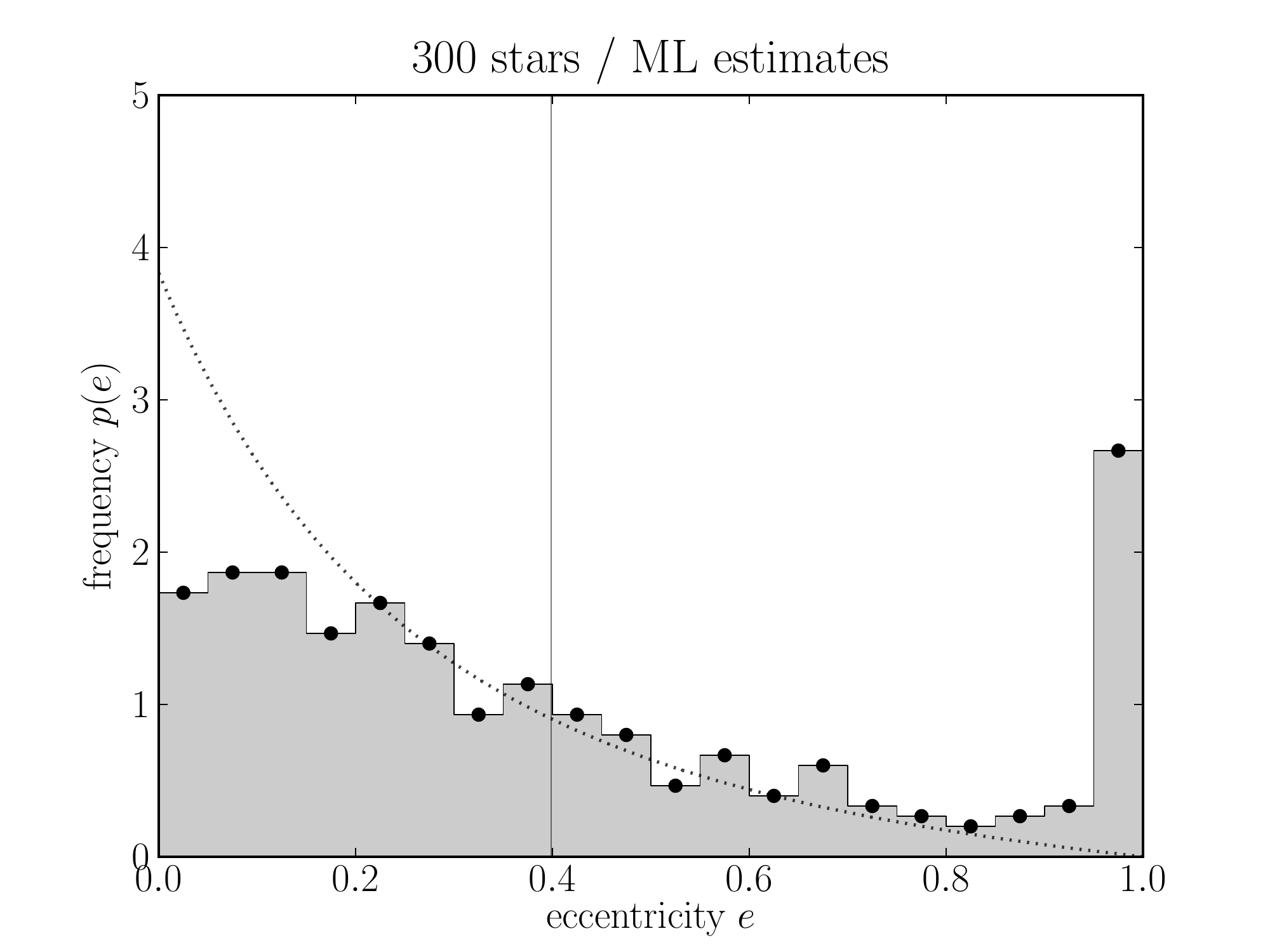}
\includegraphics[width=\figurewidth]{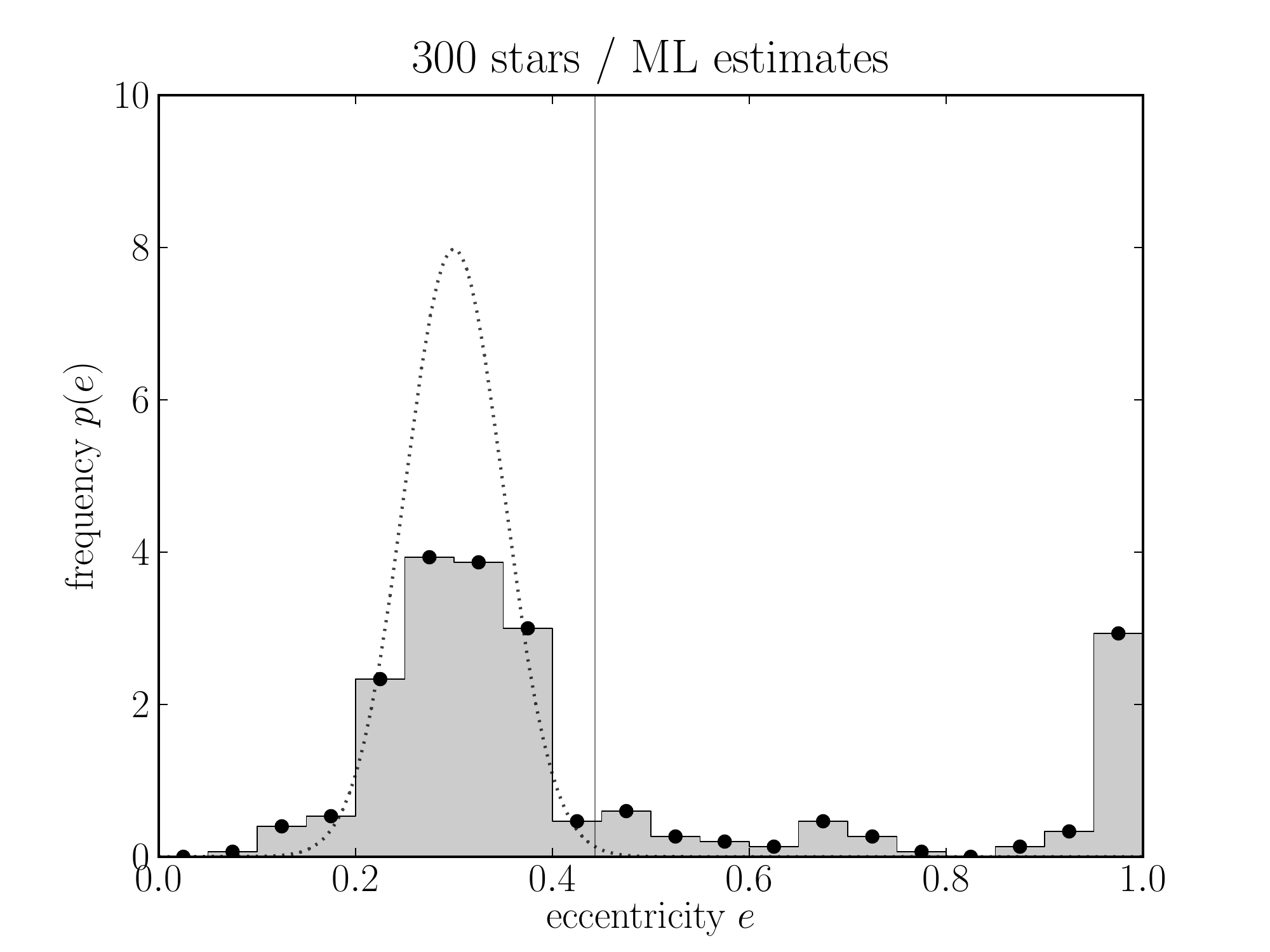}\\
\includegraphics[width=\figurewidth]{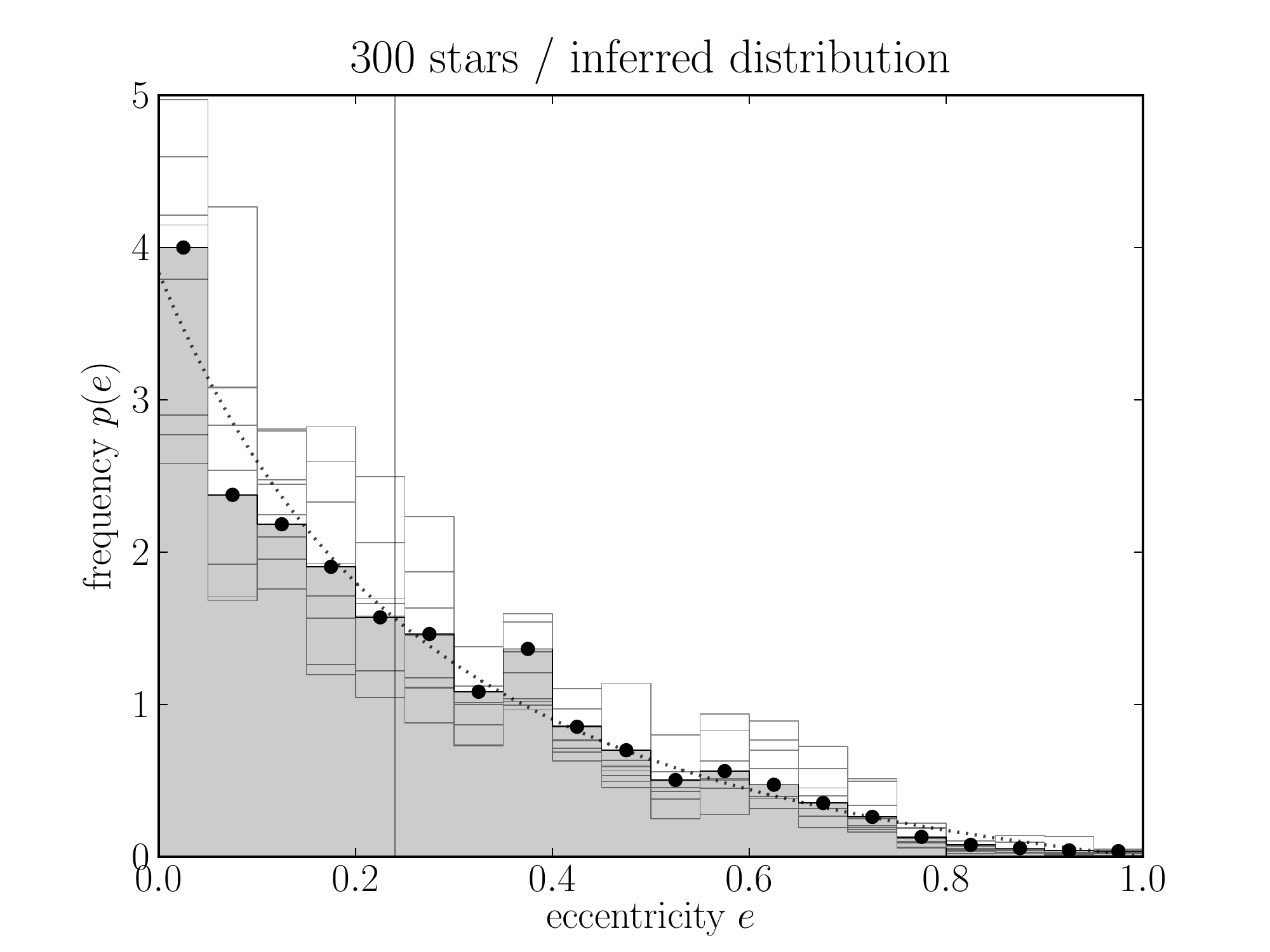}
\includegraphics[width=\figurewidth]{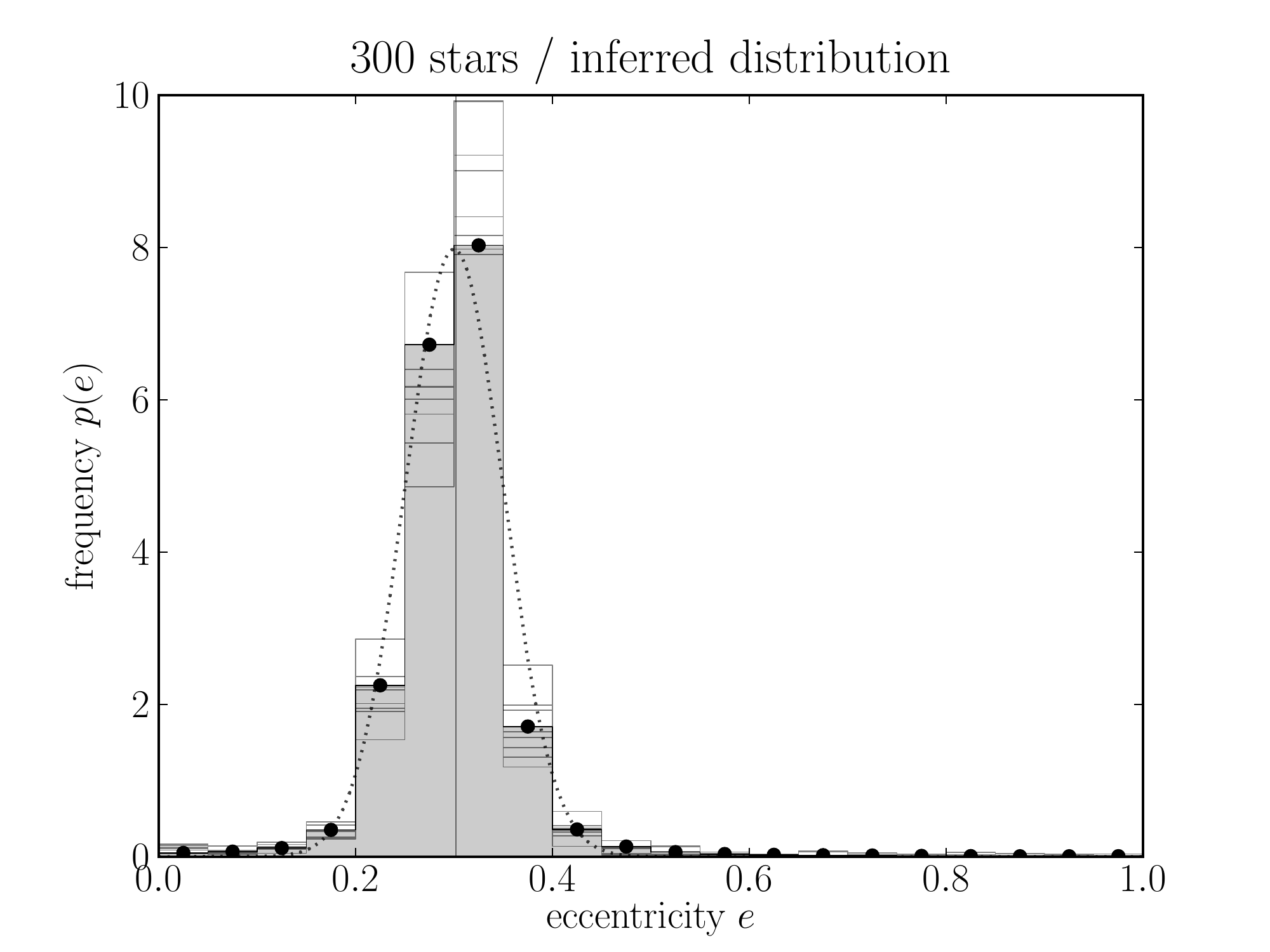}
\caption{True, maximum-likelihood, and inferred eccentricity
  distributions for two samples of 300 ersatz exoplanets. The top
  panels show---as dotted curves---the frequency distributions from
  which the true eccentricities were drawn, the ST4 distribution on
  the left, and the Gaussian distribution on the right (see text for
  details).  Superimposed is shown---as histograms---the obtained
  sampling of true eccentricities.  The vertical lines indicate the
  means of the histogrammed distributions.  The middle panels
  show---as histograms---the distributions of the maximum-likelihood
  (best-fit) eccentricities.  Again, the vertical lines show the means
  of the histogrammed distributions.  The bottom panels show a
  sampling---as a set of superimposed light histograms---of inferred
  distribution functions, drawn from the posterior PDF, and---as a
  solid line with dots---the marginalized inferred distribution (the
  mean of a $10^4$-point sampling).  The vertical line shows the mean
  of the marginalized inferred distribution (the marginalized mean).
  The dotted curves are repeated in all panels to guide the
  eye.\label{fig:300}}
\end{figure}

\begin{figure}
\includegraphics[width=\figurewidth]{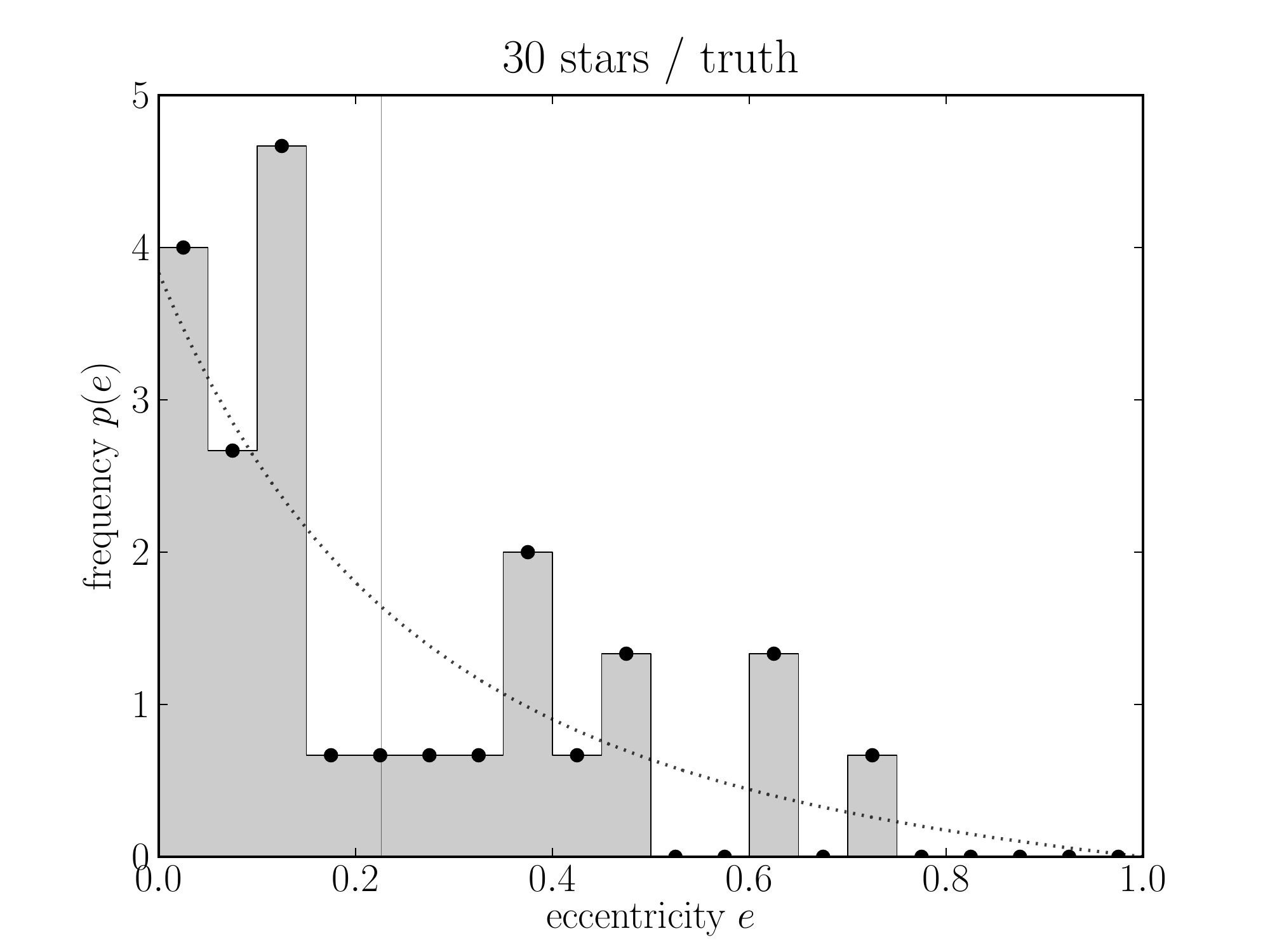}
\includegraphics[width=\figurewidth]{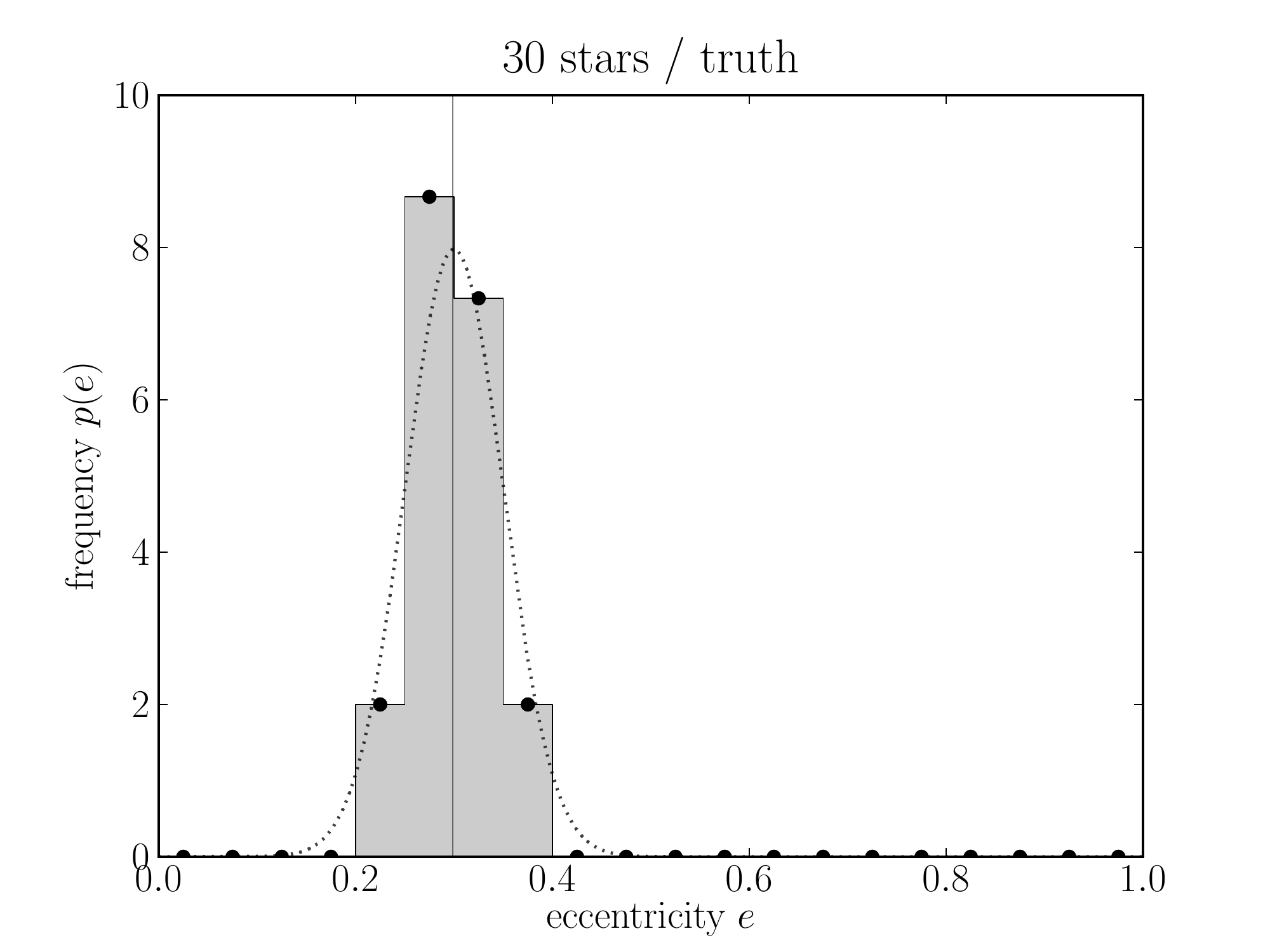}\\
\includegraphics[width=\figurewidth]{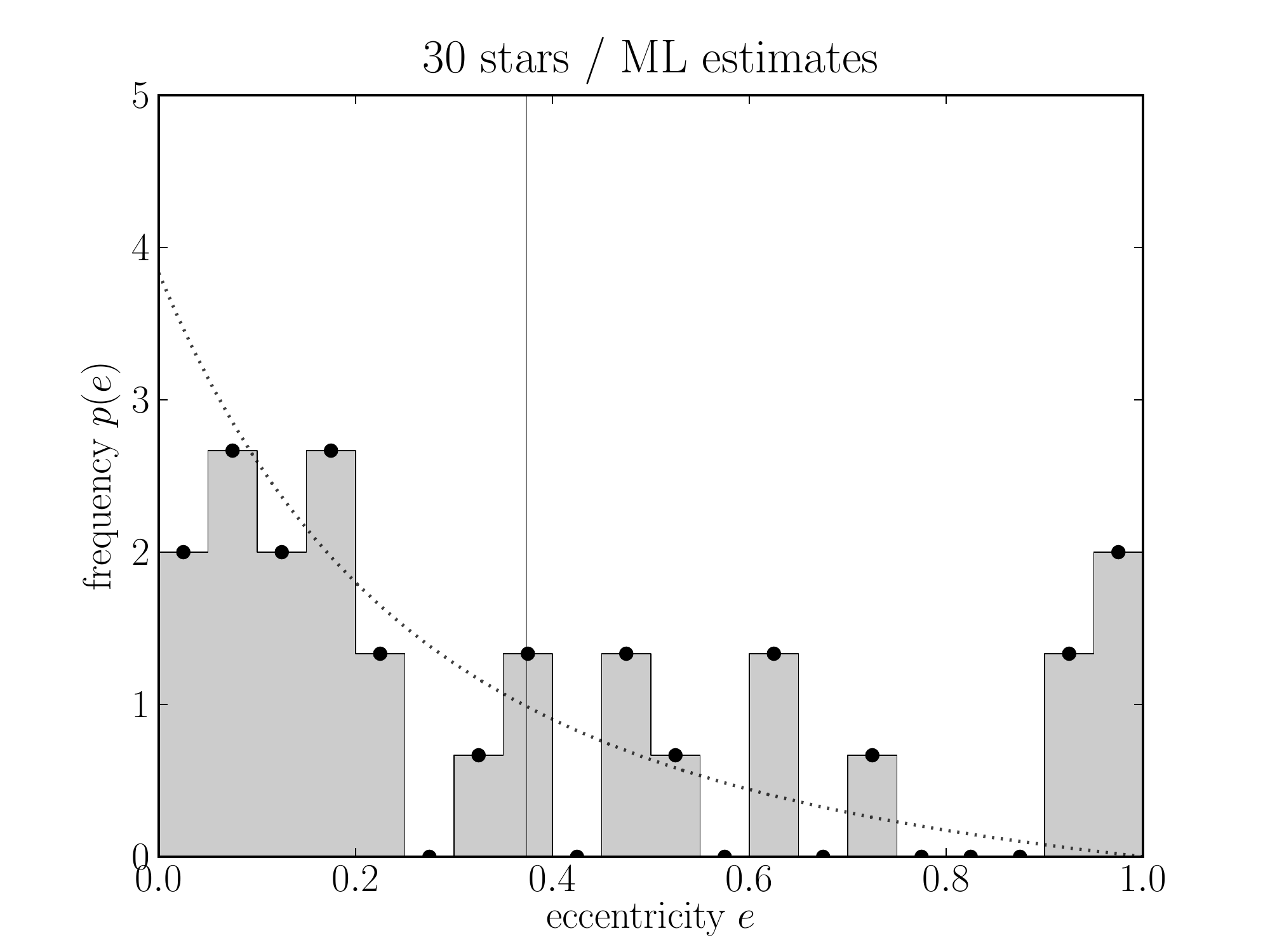}
\includegraphics[width=\figurewidth]{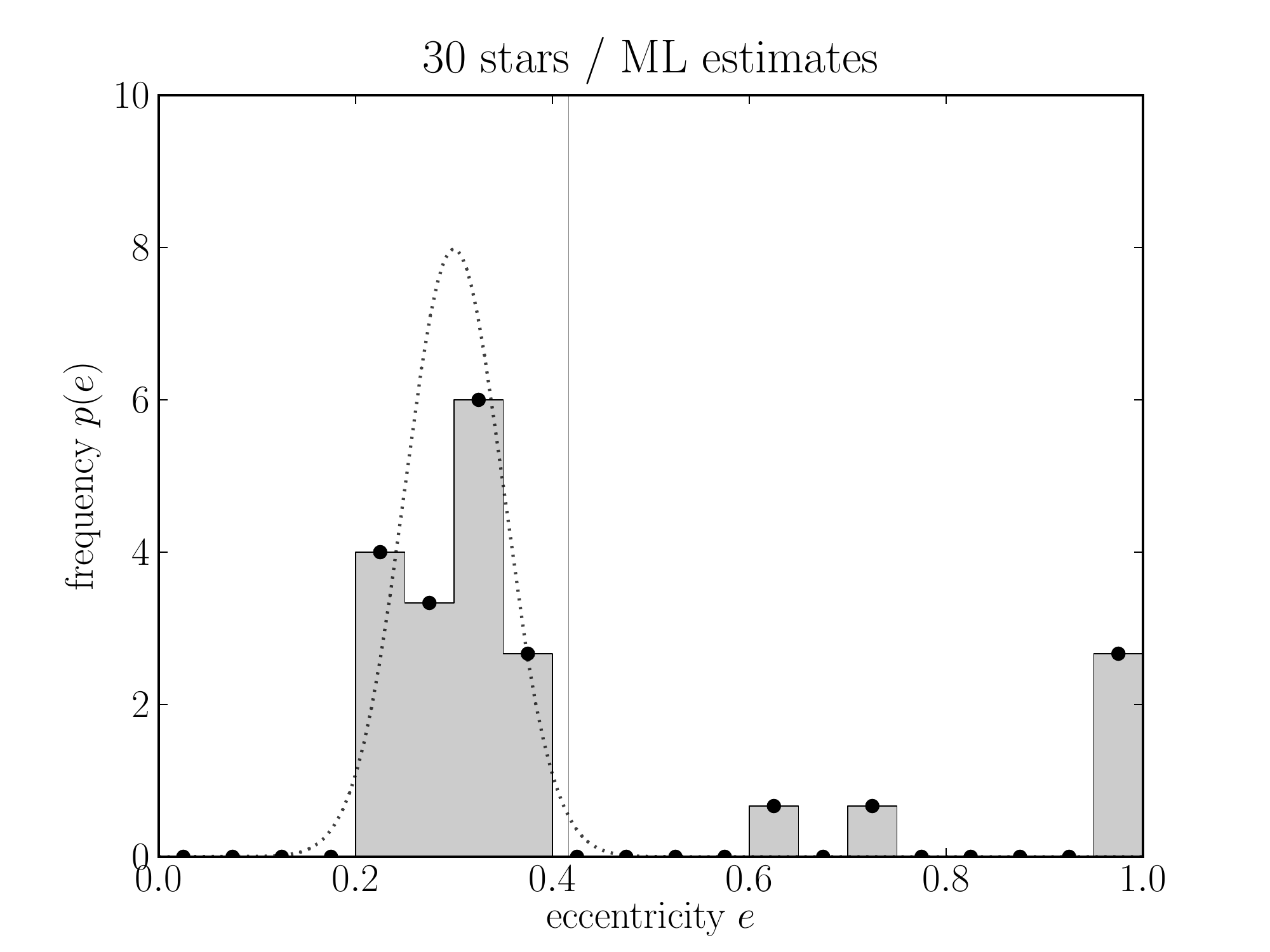}\\
\includegraphics[width=\figurewidth]{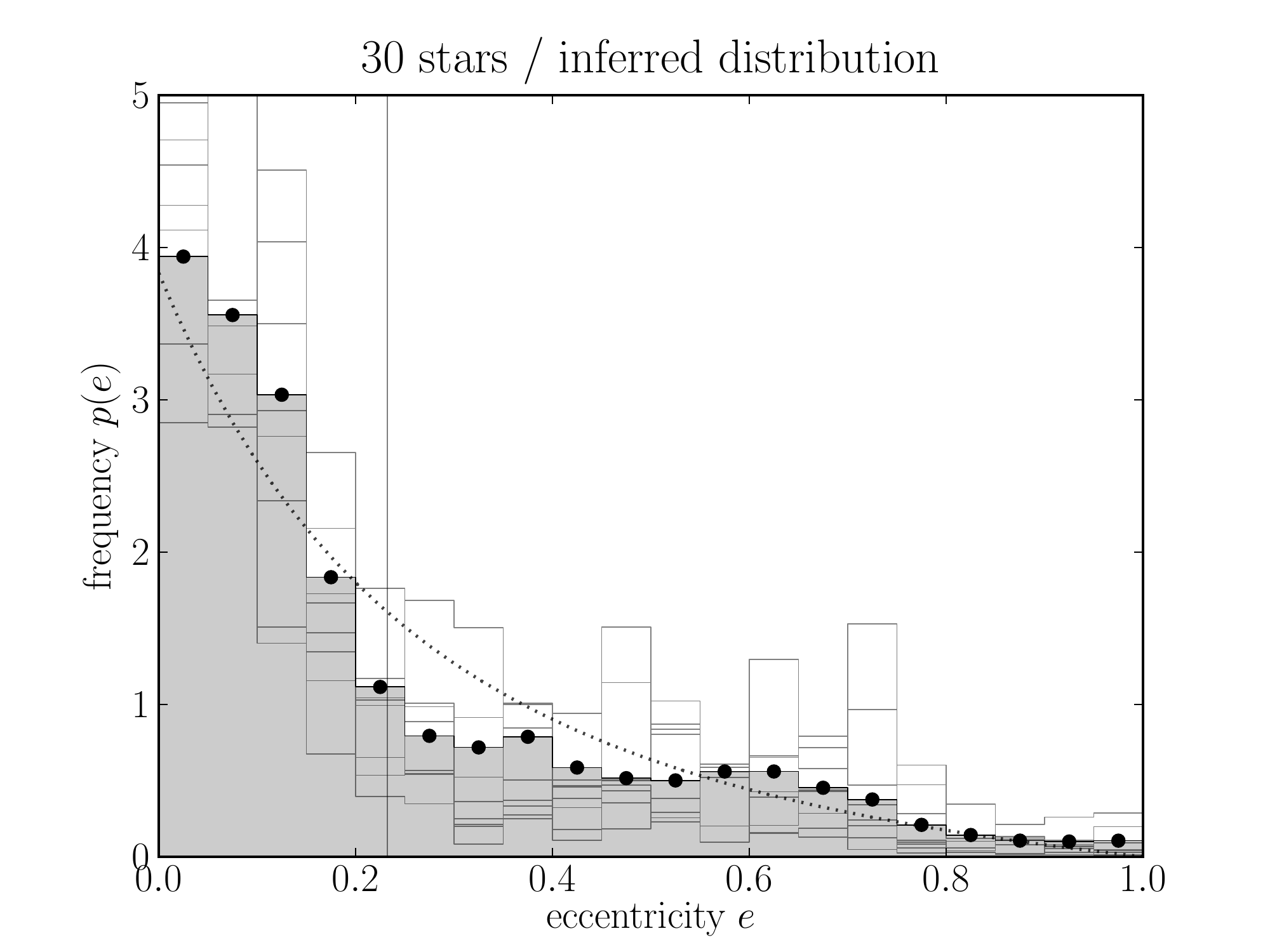}
\includegraphics[width=\figurewidth]{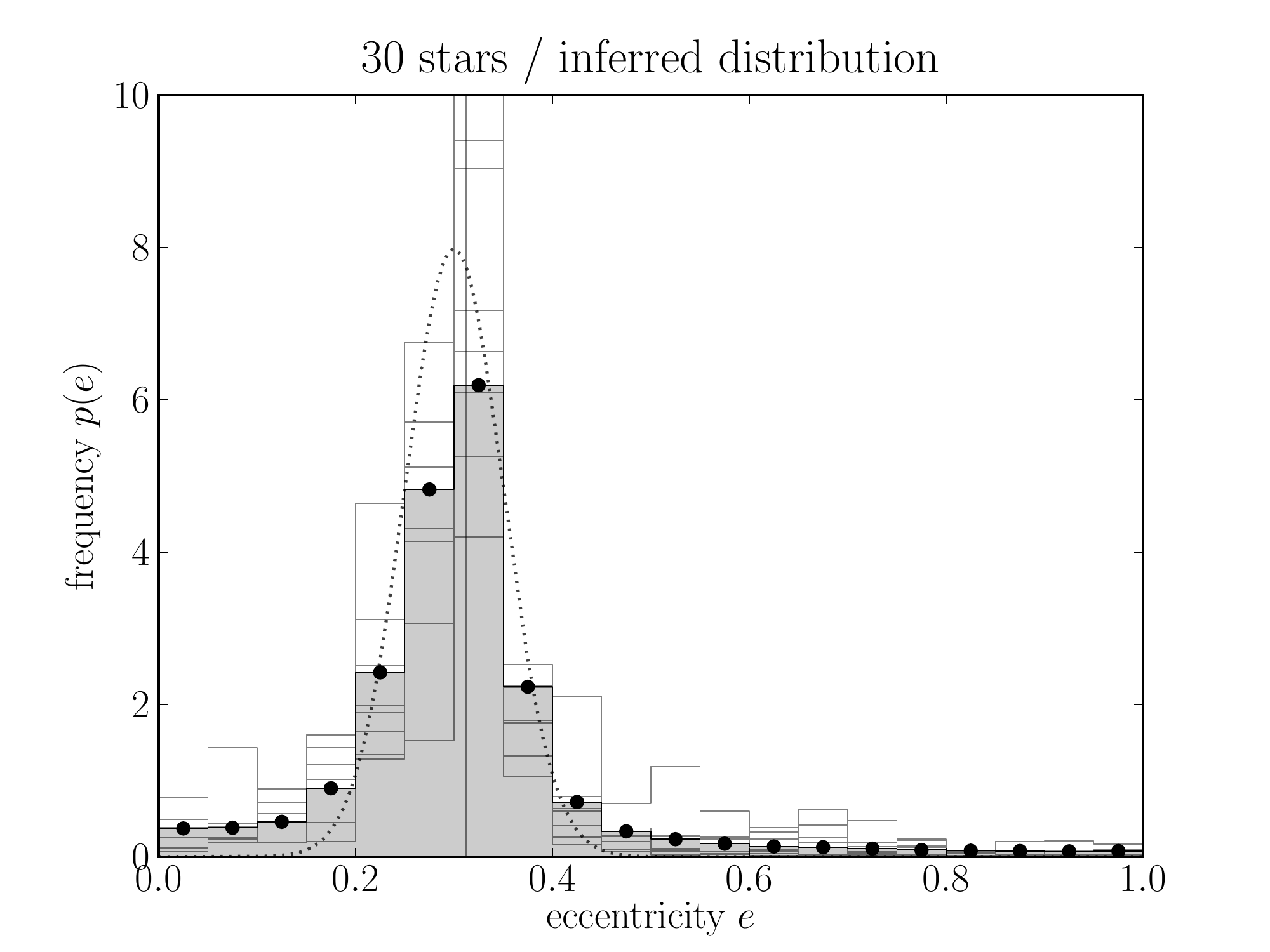}
\caption{Same as \figurename~\ref{fig:300} but for just the first 30
  ersatz exoplanets.\label{fig:30}}
\end{figure}

\begin{figure}
\includegraphics[width=\figurewidth]{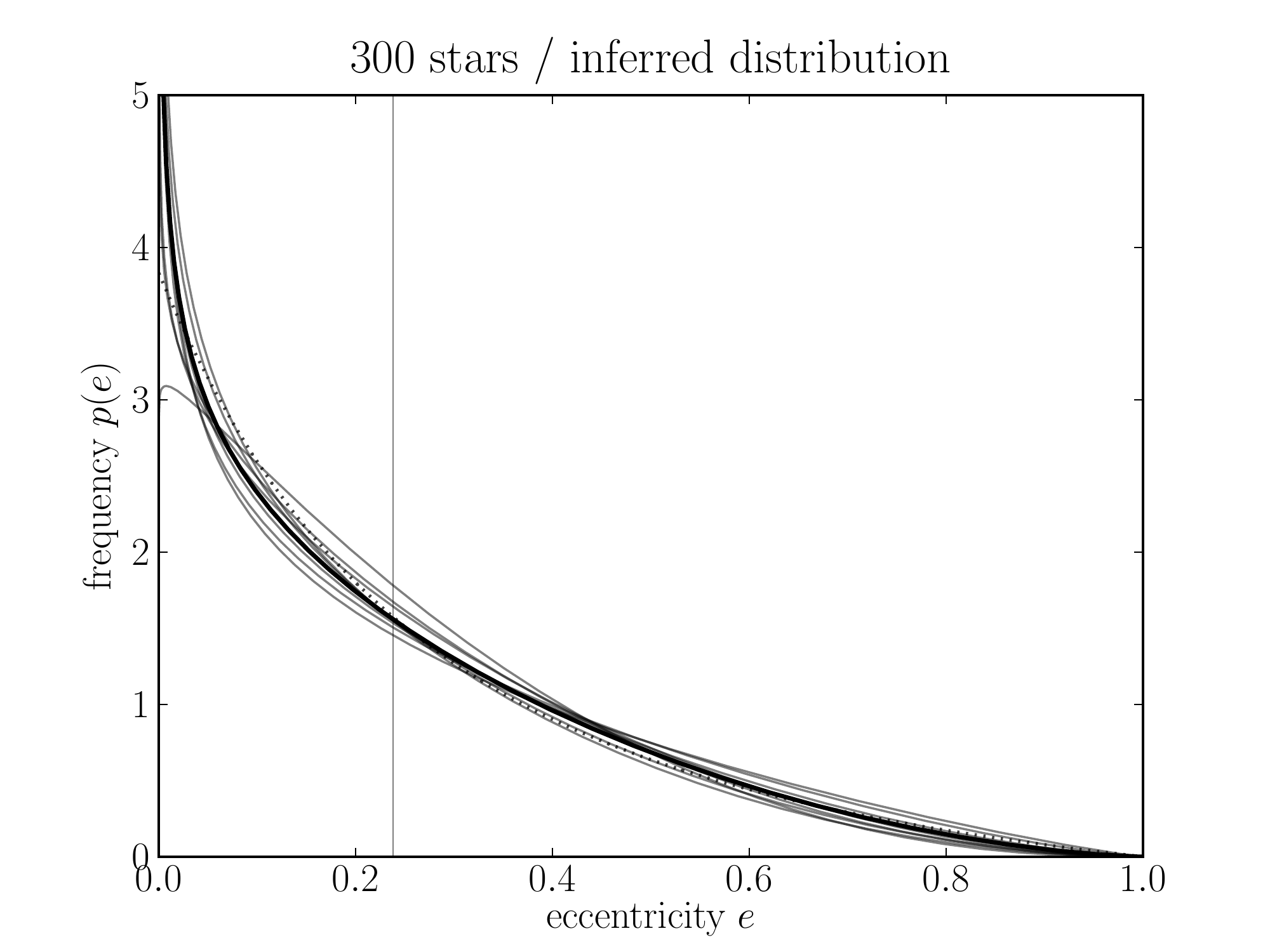}
\includegraphics[width=\figurewidth]{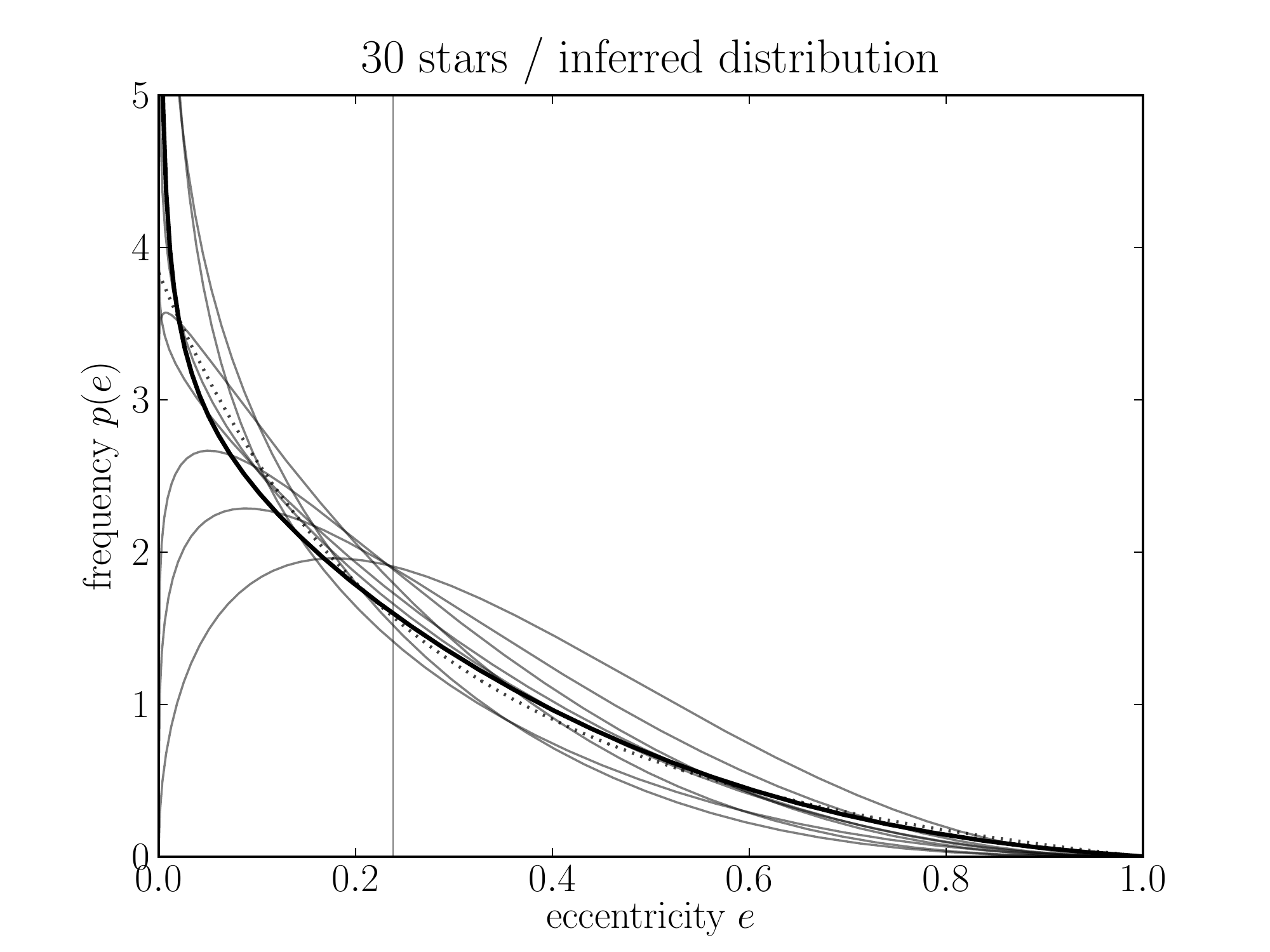}
\caption{On the left, similar to the bottom-left panel of
  \figurename~\ref{fig:300} (300 stars, ST4 distribution), but using
  the beta distribution rather than the step function for
  $\falpha(e)$.  The true ST4 distribution is shown with a dotted
  line, a sampling from the posterior PDF is shown with solid grey
  lines, and the marginalized inferred distribution (the mean of a
  $2\times10^3$-point sampling) is shown with a solid black line.  On
  the right, the same but similar to the bottom-left panel of
  \figurename~\ref{fig:30} (30 stars, ST4 distribution).\label{fig:beta}}
\end{figure}

\begin{figure}
\includegraphics[width=\figurewidth]{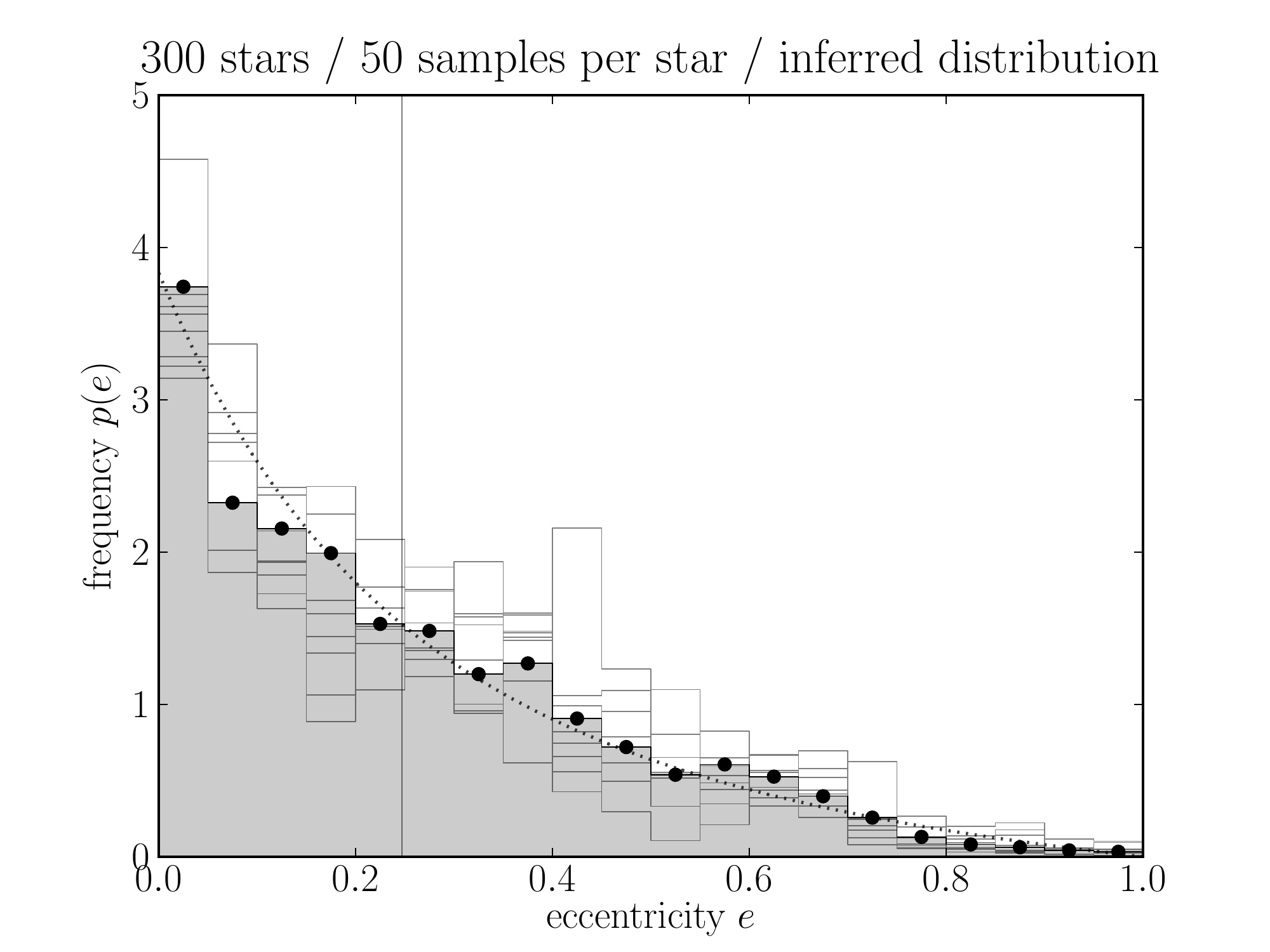}
\includegraphics[width=\figurewidth]{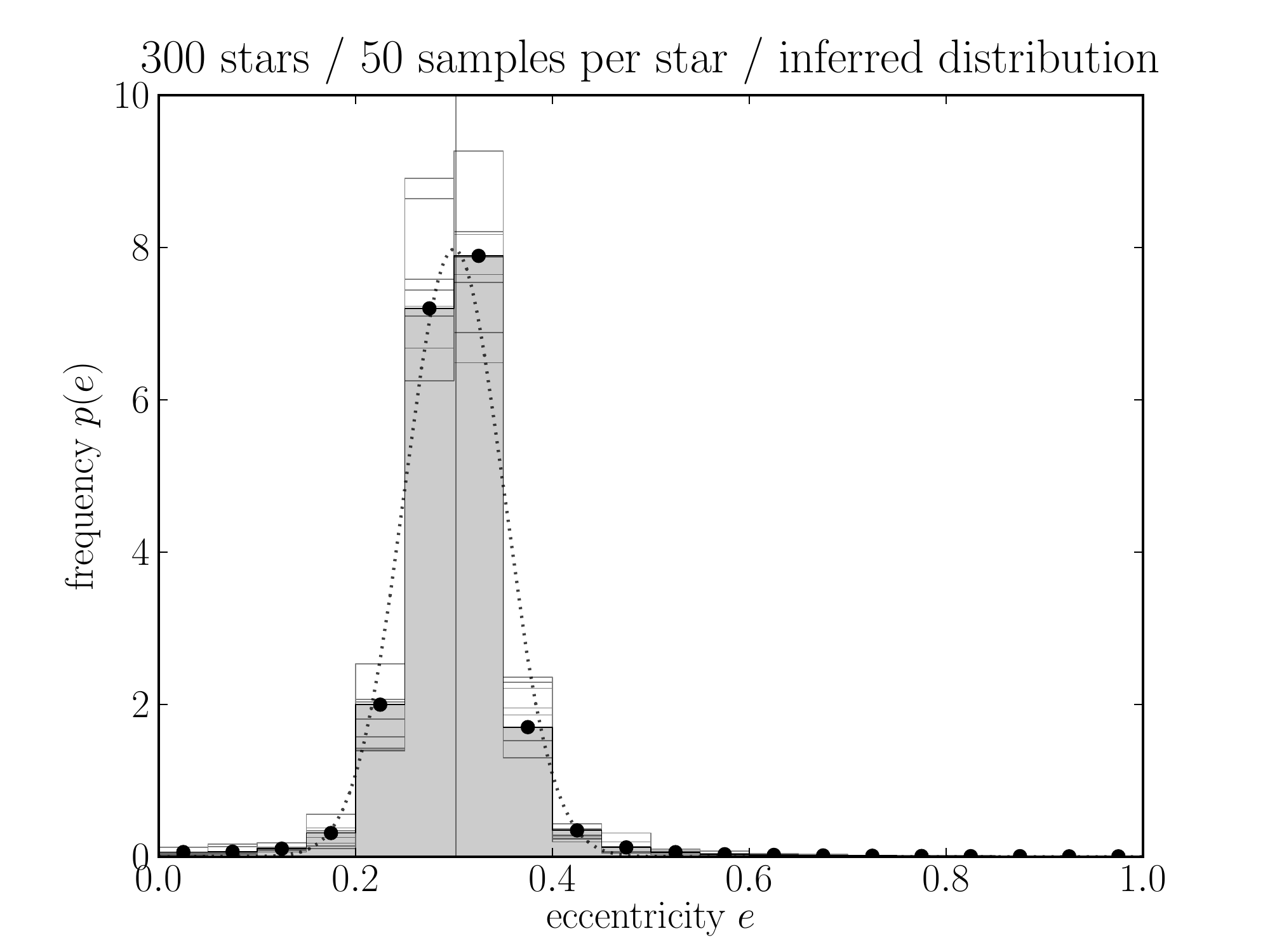}
\caption{Same as the bottom row of \figurename~\ref{fig:300} but
  thinning the individual-exoplanet posterior PDF samplings down by a
  factor of 2000, from $K=10^5$ to $K=50$ samples per
  exoplanet.\label{fig:K50}}
\end{figure}

In \figurename~\ref{fig:300} and \figurename~\ref{fig:30}, there are
many objects whose eccentricities have been over-estimated by the
maximum-likelihood method, some of them drastically.  This arises
naturally because the maximum-likelihood eccentricity tends to be
higher than the true eccentricity, but it also arises a bit
un-naturally because we have included in these figures some ersatz
systems for which the exoplanet is not detected at significant
signal-to-noise.  That is, we threw in every system, even though at
some periods, masses, and inclinations, the signal-to-noise is near or
below unity.  In most real experiments, these systems are removed
(no-one measures the eccentricity distribution for undetected
planets!).  However, this method is not thrown off significantly by
the low signal-to-noise systems.  This is a good property of a proper
probabilistic data analysis methodology: It is not moved around by the
lowest signal-to-noise objects.  Many popular methods in astrophysics
do \emph{not} have this simple property (necessary property, some
would say).  For example, in principal components analysis, the
highest-noise systems often dominate the total data variance and
therefore dominate the analysis.  In general, measurements of variance
(as measurements of distribution functions often are) can be thrown by
noisy data.  Forward models tend not to be.

In the real world, an investigator might want to infer the
eccentricity distribution, but use input from many different exoplanet
observers, each of whom might have sampled their exoplanets with
different uninformative priors $p_0(e)$.  This is not a problem; in
the importance sampling of \equationname~(\ref{eq:like}), the sum for
each object $n$ should make use of the uninformative prior $p_0(e)$
used on object $n$.  Also in the real world, an investigator might
want to infer the eccentricity distribution using a mixture of
measurements, some of which have been delivered as samplings, and some
of which have been delivered just as maximum-likelihood estimates.
Unfortunately these latter---maximum-likelihood estimates or any other
single-point estimates---are nearly useless for modeling.

The method presented here is a baby step towards a hierarchical
Bayesian method.  The method would be fully hierarchical if we went
back to the objects and re-sampled them using the inferred prior, or,
even better, inferred the prior and performed the
individual-object samplings simultaneously.  That is, this method
becomes fully hierarchical when the inferred distribution function is
used to improve the individual-object estimates. When the measurements
are given as likelihood functions, this should be the preferred
method.

Along those lines, we could have performed a less simple but faster
hierarchical sampling: Rather than marginalizing over the individual
eccentricities by summing over a (potentially large) number of
eccentricity samples for each exoplanet, we can use the provided
samplings of the individual eccentricities as the basis of an approach
that samples both the parameters of the eccentricity distribution and
the true eccentricities of the planets. This approach avoids the sum
in \equationname~(\ref{eq:like}).  It also returns updated posterior
samples of the individual eccentricies using the better prior---the
prior inferred from the population of planets.

\begin{figure}
\includegraphics[width=\figurewidth]{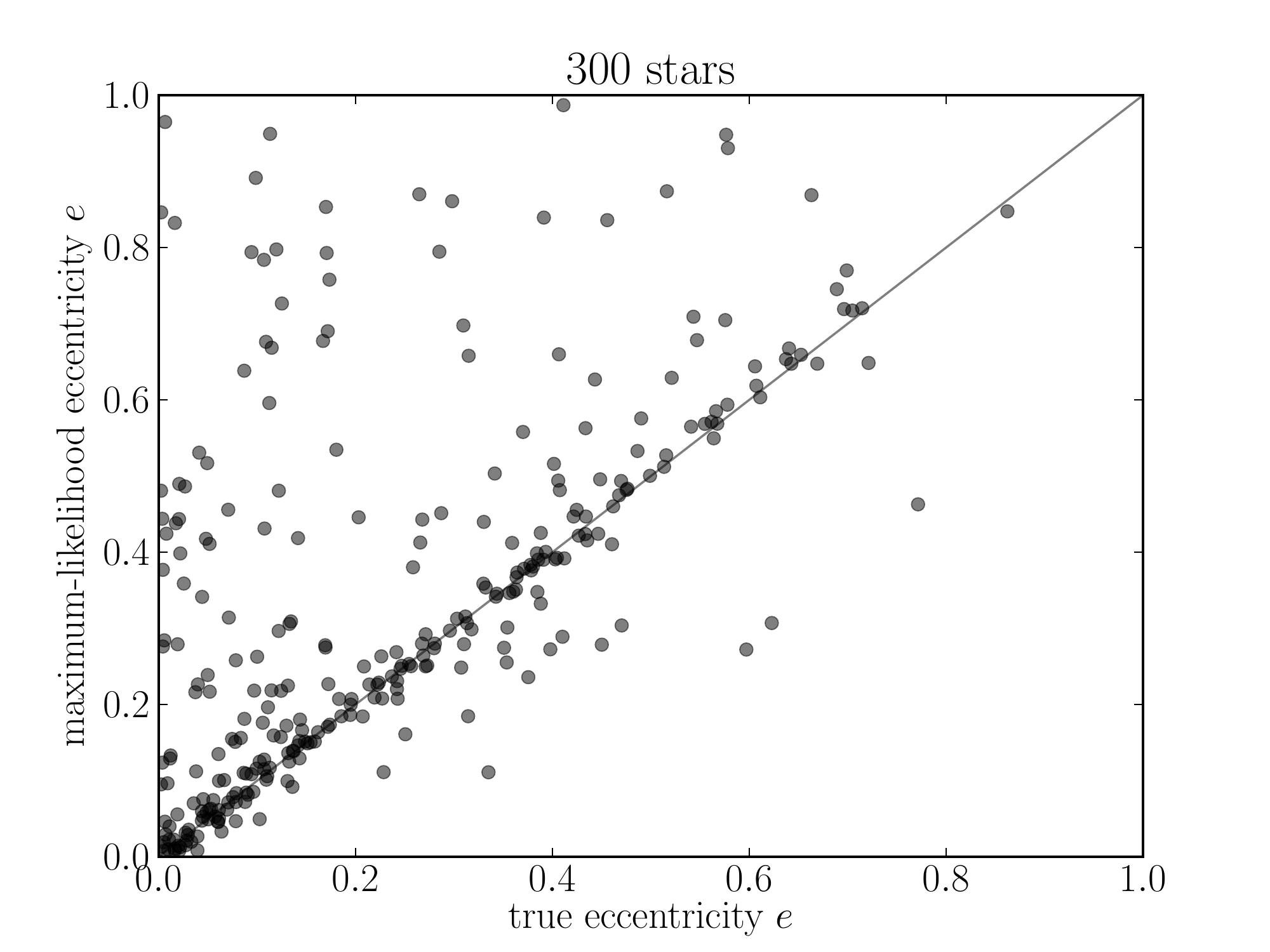}
\includegraphics[width=\figurewidth]{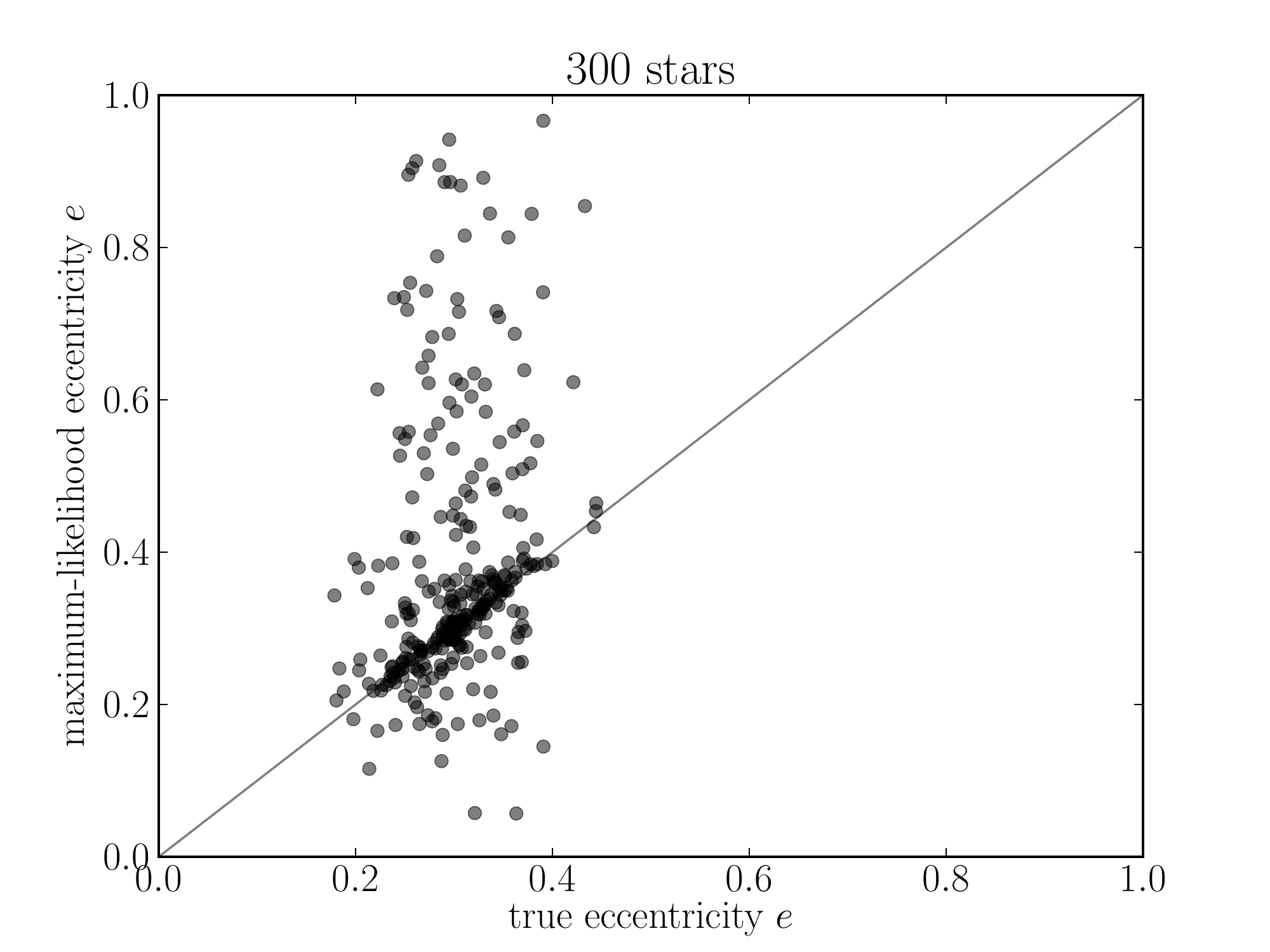}\\
\includegraphics[width=\figurewidth]{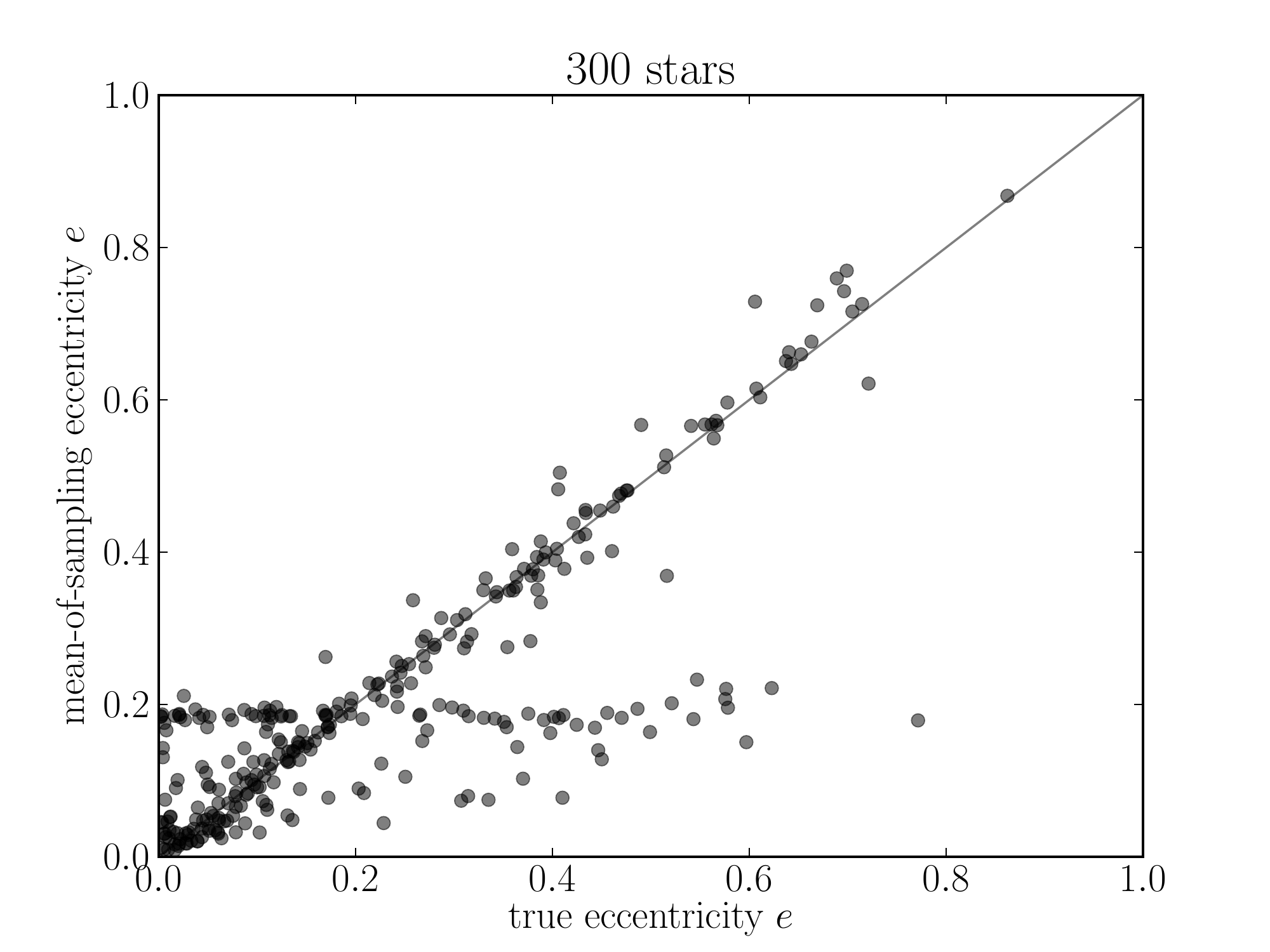}
\includegraphics[width=\figurewidth]{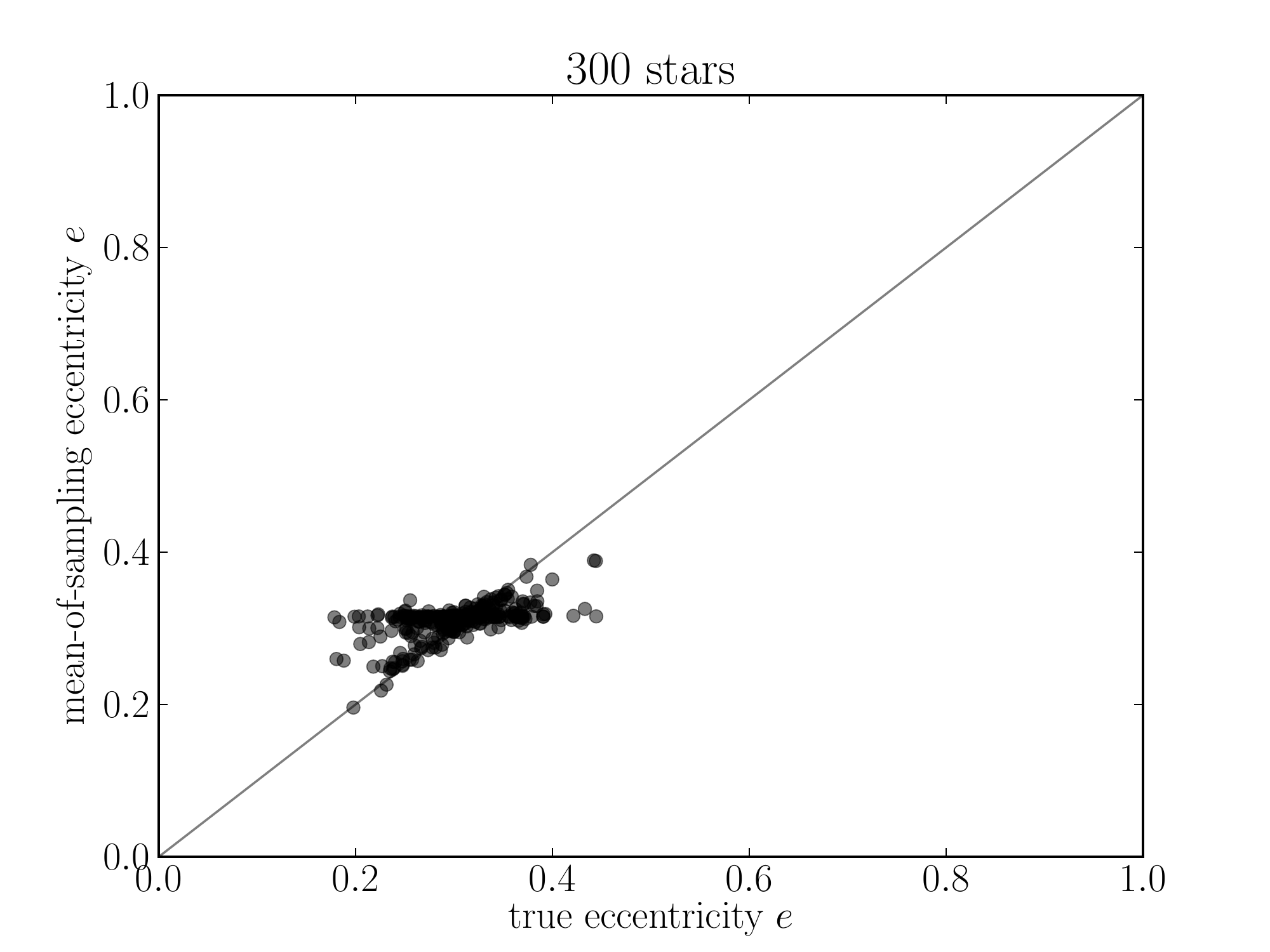}
\caption{Improving point estimates hierarchically.  The top row shows
  a comparison of true eccentricities (for our ersatz exoplanets) with
  the maximum-likelihood estimates, for the ST4 (left) and Gaussian
  (right) true distributions.  The bottom row shows the comparison of
  mean-of-sampling eccentricity estimates (see text), made using as
  prior PDFs on the eccentricity the (highly informative) marginalized
  inferred distributions of \figurename~\ref{fig:300}.  The horizontal
  line near 0.2 in the left figure and near 0.3 in the right figure
  comes from very low signal-to-noise systems (signal-to-noise at or
  below unity) for which the mean-of-sampling estimate ends up being
  very close to the mean of the prior PDF; this is the only reasonable
  mean estimate when the data are not informative.\label{fig:compare}}
\end{figure}

Briefly, this approach entails writing down the joint posterior
probability of the individual eccentricies and the parameters of the
eccentricity distribution. Using Bayes's theorem, we can write this
joint distribution as
\begin{eqnarray}\displaystyle\label{eq:hierarchicaljoint}
p(\{e_{n}\}, \alphavec | \allv) &\propto& p(\allv |\{e_{n}\})\,p(\{e_n\} | \alphavec)\,p(\alphavec)\,.
\end{eqnarray}
By sampling from this posterior distribution, we simultaneously obtain
samples of the eccentricities of the individual planets and of the
parameters of the eccentricity distribution. These eccentricity
samples will then have used the better eccentricity prior
parameterized by $\alphavec$, rather than the uninformative original
prior. For planets detected at low signal-to-noise, this leads to more
realistic estimates of the eccentricity (see
\figurename~\ref{fig:compare}).

We can re-write \equationname~(\ref{eq:hierarchicaljoint}) as
\begin{eqnarray}\displaystyle
p(\{e_n\}, \alphavec | \allv) 
&\propto& \left[p(\allv | \{e_{n}\})\,p_0(\{e_n\})\right]  \,\frac{p(\{e_n\} | \alphavec)}{p_0(\{e_n\})}\,p(\alphavec)
\quad.
\end{eqnarray}
The provided eccentricity chains give us samples of the distribution
in square brackets. We can re-use these samples in a manner similar
to that in \equationname~(\ref{eq:like}) to sample from the joint
distribution of $\{e_n\}$ and $\alphavec$.  Starting from initial
$(\{e_n\}^{(i)},\alphavec^{(i)})$, first
\textsl{(1)}~Metropolis-sample $\{e_n\}^{(i+1)}$ from $p(\{e_n\} |
\allv,\alphavec^{(i)})$ by sampling from
\begin{eqnarray}\displaystyle
p(\allv | \{e_{n}\})\,p_0(\{e_n\})
\quad ;
\end{eqnarray}
which can be done by just picking a random sample from the given
eccentricity chains for the individual planets---and using the
Metropolis-Hastings acceptance probability on
\begin{eqnarray}\displaystyle
\frac{p(\{e_n\}^{(i+1)}|\alphavec^{(i)})\,p_0(\{e_n\}^{(i)}|\alphavec^{(i)})}
     {p(\{e_n\}^{(i)} \alphavec^{(i)})\,p_0(\{e_n\}^{(i+1)}|\alphavec^{(i)})}
\quad .
\end{eqnarray}
Using this acceptance probability ensures that the samples are
``importance-resampled'' according to the new prior parameterized by
$\alphavec$. Then, \textsl{(2)}~sample $\alphavec^{(i+1)}$ from
$p(\alphavec | \{e_n\}^{(i+1)})$ using any MCMC sampler. The resulting
sampling of $\alphavec$ can then be used in exactly the same way as in
\sectionname~\ref{sec:experiments}.

Just to demonstrate the power of the hierarchical approach, in
\figurename~\ref{fig:compare} we show a comparison of the
maximum-likelihood eccentricity estimates to the mean-of-sampling
estimates made with the inferred prior.  That is, we take the
marginalized inferred distribution $\left<\falpha(e)\right>$,
re-weight the eccentricity samples with this inferred distribution,
and produce as a point estimate for each system $n$ the mean of the
re-weighted sampling.  These mean-of-sampling estimates, made with a
correctly informative prior, are---not surprisingly---better than the
maximum-likelihood estimates.

In general, hierarchical inference must become a standard tool in
astrophysics going forward: Our science is fundamentally statistical,
and the objects of greatest interest are measured always at the limits
of instrumental sensitivity.  Hierarchical methods are the right tools
for these jobs.

\acknowledgements It is a pleasure to thank Brendon Brewer (UCSB),
Fengji Hou (NYU), Dustin Lang (Princeton), Iain Murray (Edinburgh),
Daniel Mortlock (Imperial), Hans-Walter Rix (MPIA), Sam Roweis
(deceased), and Christian Schwab (Yale) for valuable discussions, and
an anonymous referee for useful feedback.  DWH and JB acknowledge
financial support from NASA (grants NNX08AJ48G), the NSF (grant
AST-0908357), and a Research Fellowship of the Humboldt Foundation.
ADM acknowledges financial support from NASA (grants NNX08AJ28G and
GO9-0114).  This research made use of the \project{SAO/NASA
  Astrophysics Data System}, the \project{Python} programming
language, and open-source software in the \project{numpy} and
\project{matplotlib} projects.  All code used in this project is
available from DWH upon request.

\clearpage

\clearpage

\clearpage

\clearpage

\clearpage

\clearpage

\end{document}